\documentclass[12pt]{article}
\usepackage{ssi}
\usepackage{epsf}

\begin{document}

\title{Measurement of $CP$ Asymmetries at Belle}

\author{William Trischuk\thanks{With support from Princeton University and the Japanese Accelerator Laboratory, KEK.}\\ 
Department of Physics \\
University of Toronto, Toronto, Ontario,  \\
M5S 1A7, Canada\\[0.4cm]
Representing the Belle Collaboration
}

\maketitle
\begin{abstract}

The Belle experiment at the KEK $B$ factory has collected 93 fb$^{-1}$
of electron positron collisions at $\sqrt s = 10.6$ GeV. This has produced
a sample of 85 million $B {\bar B}$ meson pairs that can be used to study
$CP$ violation in rare (and not so rare) $B$ decay modes. Here I report on
a measurement of indirect $CP$ violation in the decay 
$B^0 \rightarrow J/\psi K^0_S$, as well as time dependent $CP$ asymmetries 
in rarer modes such as $B^0 \rightarrow \pi^+ \pi^-, \eta' K_S^0$
and $\phi K^0_S$. 
I summarise the prospects for improving the precision 
on these and related measurements.

\end{abstract}

\section{$CP$ Violation in $B$ Decay}

When $CP$ violation was first 
observed in neutral kaon decay, in the early 1960s, it shook
the foundations of particle physics. It had previously been assumed
that the combination of charge conjugation and a parity transformation
left all known particle interactions invariant, despite the fact
that weak interactions violated parity alone.
Over the following three decades $CP$ violation
in the $K^0$ system was measured with ever increasing precision in an attempt 
to pin down its source. In the 1970s Kobyashi and Maskawa~\cite{km} showed 
that the Standard Model could
accommodate $CP$ violation in a three quark weak mixing
matrix, $V_{\rm CKM}$. The  
single non-trivial phase in such a $3 \times 3$ unitary matrix
could explain the small effect first seen
in $K^0$ meson decay, where $CP$ violation
was observed at the $10^{-3}$ level. It was suggested~\cite{Sanda},
in the early 1980s, that the comparable amplitudes for
the direct decay of $B^0$ mesons into $CP$ eigenstates
and the mixing of $B^0 {\bar B^0}$ mesons would make neutral $B$
meson decay an ideal place to observe large indirect
$CP$ violating effects.

\begin{figure}
\begin{center}
\vspace*{-0.6in}
\mbox{\epsfxsize 10cm \epsfbox{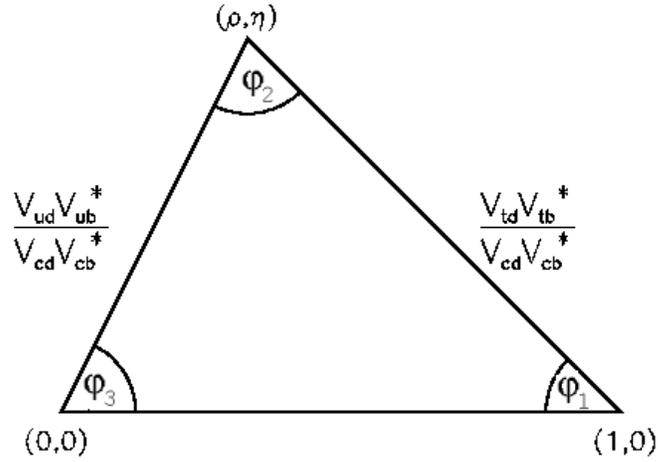}}
\vspace*{-0.3in}
{\caption [The Unitarity Triangle]
{\label{unitarity}
Unitarity of the CKM matrix relates its different
elements. The study of $CP$ violation in $B$ meson
decay involves the CKM elements depicted
in this triangular relationship. }}
\end{center}
\end{figure}

One way to understand the magnitude of $CP$ violation
predicted by the CKM model, in neutral $B$ meson decay, is to
consider the unitarity relation between the first and third
columns of $V_{\rm CKM}$:
$$ V_{tb}^*V_{td} + V_{cb}^*V_{cd} + V_{ub}^*V_{ud} = 0. $$
While each term in this expression is relatively
small (${\cal O}(\sin^3\theta_C)$), they are all the same
size. When plotted in the complex plane
(see fig.~\ref{unitarity}) one expects 
significant angles at each apex because the sides of
the triangle have similar lengths. In neutral kaon decay
the corresponding unitarity triangle has two sides that
are much larger than the third -- making the decay
rates for $K^0$ mesons much larger, but making the angles,
and hence the observable phases small and
more challenging to measure.

In $B$ meson decay $CP$ violating phases are most
readily observed through the indirect mixing of two
amplitudes. Given a $CP$ eigenstate
accessible to both $B^0$ and ${\bar B^0}$ decays --
such as $J/\psi K^0_S$ -- one can observe
the interference between the direct decay amplitude for:
$$ B^0 ({\bar B^0}) \rightarrow J/\psi K_S^0; $$
and the amplitude for the same decay preceded by $B^0$ mixing:
$$ B^0 ({\bar B^0}) \rightarrow {\bar B^0} (B^0) \rightarrow J/\psi K_S^0. $$
When one includes the relative phase between these two
amplitudes we get an expression for the time
dependent $CP$ asymmetry:

\begin{center}
\begin{tabular}{ccl}
$A_{CP}(\Delta t)$ & $\equiv$ & ${ { {dN \over dt} (\bar{B^0} \rightarrow J/\psi K_{\xi}) - 
               {dN \over dt} (B^0 \rightarrow J/\psi K_{\xi}) } \over 
               { {dN \over dt} (\bar{B^0} \rightarrow J/\psi K_{\xi}) + 
               {dN \over dt} (B^0 \rightarrow J/\psi K_{\xi} } }$, \\ 
            & = & - $\xi_K \sin~2\phi_1 \> \sin~\Delta m \Delta t. \qquad \qquad [1]$
\label{asym}
\end{tabular}
\end{center}
Where the asymmetry in the decay rate between $B^0$ and ${\bar B^0}$ mesons
is proportional to the $B^0$ mixing rate ($\sin \Delta m \Delta t$), the
$CP$ eigenvalue of the final state, $\xi_K$, ($\xi_K = -1$ for $J/\psi K^0_S$
and +1 for $J/\psi K^0_L$) and $\sin 2 \phi_1$,
the angle at the lower right apex of the unitarity 
triangle shown in fig.~\ref{unitarity}.

\section{The KEK-B Collider and Belle Detector}

The main experimental challenge in measuring $CP$ violation in $B^0$ meson
decay lies in the fact that $B^0$ mesons decay much more
quickly than $K^0$ mesons, having proper flight distances of
fractions of a millimeter. Furthermore
the decay rate to experimentally accessible 
$CP$ eigenstates are much
smaller, on the order of ${\cal O} (10^{-4})$ for $B^0$ mesons; 
while essentially
100\% of $K^0$ meson decays are to identifiable $CP$
eigenstates. While the very much larger $CP$ violation in $B^0$
decay makes up for some of this it remained a significant
experimental challenge to observe $CP$ violation in $B^0$ meson decay.

\begin{figure}
\begin{center}
\vspace*{-0.6in}
\mbox{\epsfxsize 13cm \epsfbox{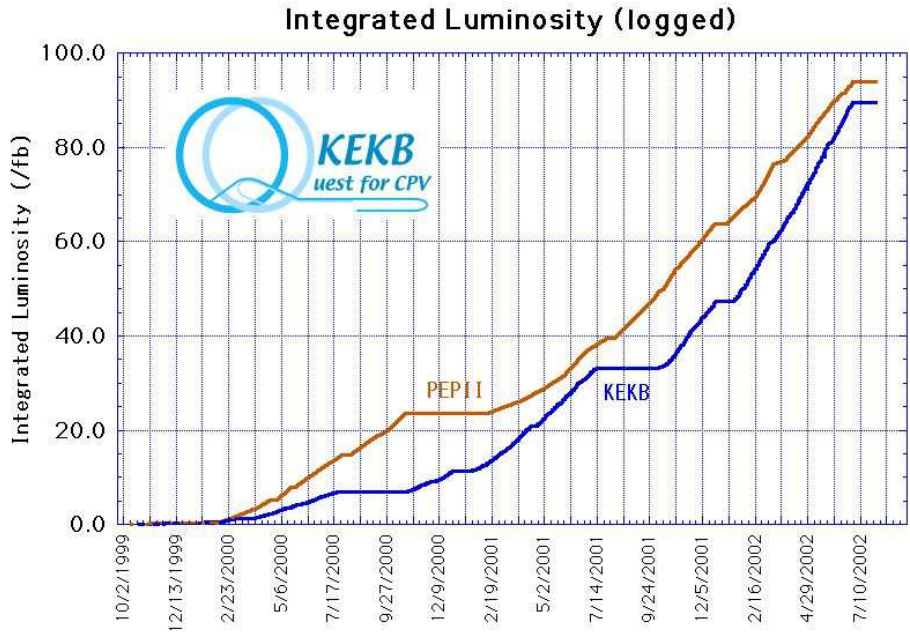}}
\vspace*{-0.2in}
{\caption [The Belle/KEKB data sample]
{\label{luminosity}
The Belle experiment has been collecting data over
the last two years. This plot shows the 
integral luminosity over that period (lower line)
compared to the luminosity collected by the
BaBar experiment at PEPII (upper line). The data
samples are similar at this point. }}
\end{center}
\end{figure}

The key ingredient is having
a high luminosity source of $B$ mesons -- a $B$ factory. Two
dedicated machines were built in the late 1990s to address
this. The KEK-B $e^+e^-$ accelerator complex
collides beams
of 8 GeV electrons and 3.5 GeV positrons, to produce
$\Upsilon(4S)$ mesons at $\sqrt s = 10.6$ GeV. These, in turn, decay
into $B {\bar B}$ meson pairs. Using 
an 11 mrad crossing angle KEK-B 
is able to collide beams with currents in excess of 1 Ampere, 
with tolerable backgrounds and luminosities approaching $10^{34}$
per cm$^2$ each second. An integrated luminosity comparison between
the KEK-B factory and its competitor, PEP-II at SLAC, is
shown in fig.~\ref{luminosity}. While the two machines
have delivered similar integrated luminosities since
they came online two years ago, the recent instantaneous
luminosities (the slope of the curve) in the KEK machine 
bodes well for upcoming data-taking.

\begin{figure}
\begin{center}
\vspace*{-0.6in}
\mbox{\epsfxsize 13cm \epsfbox{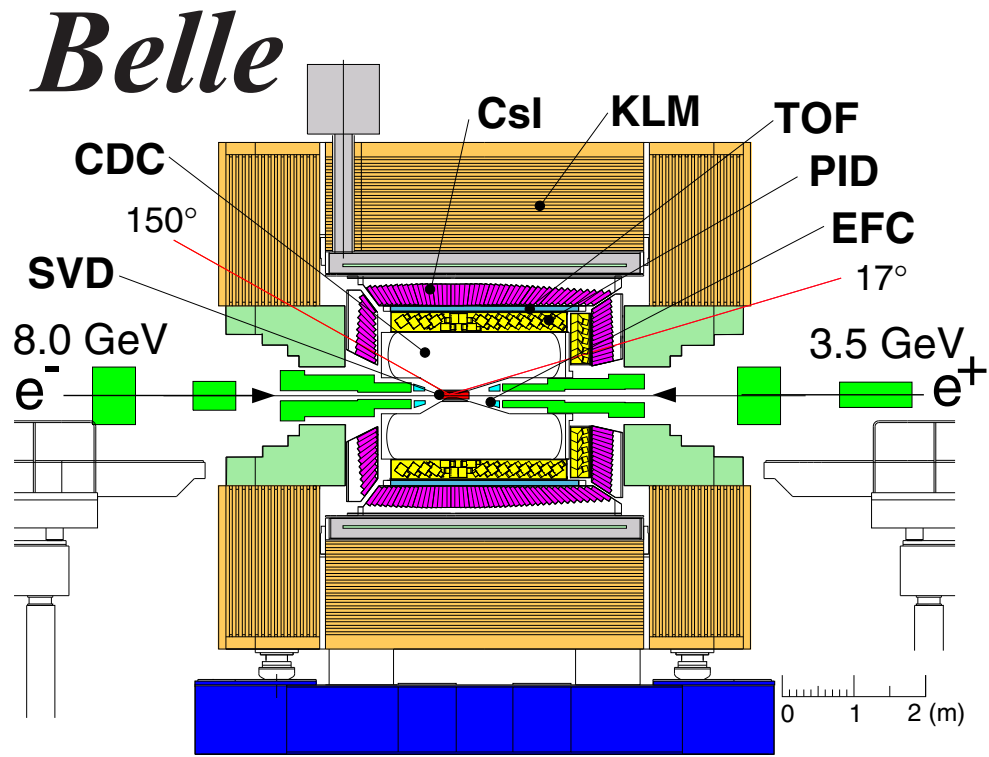}}
{\caption [The Belle detector]
{\label{detector}
An elevation view of the Belle experiment. }}
\end{center}
\end{figure}

The colliding beam energies are asymmetric so that the $\Upsilon(4S)$
meson produced is boosted along the beam direction. This results in the
daughter $B$ mesons also being boosted, with $\beta \gamma \approx 0.45$,
separating the two $B$ meson decay
vertices in the detector. The boost introduces some asymmetry
in the design of the experiment. A cross-section of
the cylindrical Belle detector is shown in fig.~\ref{detector}.
The detector elements are arranged asymmetrically
around the interaction point to increase the acceptance for
$B$ meson decay products.

Working outwards from the collision point $B$ meson decay
products encounter a three layer silicon vertex detector (SVD)
that measures charged particle trajectories with 55 $\mu$m
precision (at 1 GeV/c) and the separation of 
$b$ decay vertices with a precision of 100 $\mu$m
in $z$. They next
pass through a Helium filled drift chamber (CDC) that measures
track momenta with a precision of 
$\sigma_p /p = (0.2p \oplus 0.3 ) \%$ ($p$ measured
in GeV/c) 
providing excellent mass resolution for $B$ decays into charged daughter
tracks. This is followed by a Cesium-Iodide crystal 
calorimeter (CsI) that has better than 2\% energy resolution for
1 GeV photons. Belle's particle identification system
includes an aero-gel Cerenkov counter system (PID) and a time of flight
system (TOF) that can distinguish kaons from pions up to 
3.5 GeV/c with 90\% efficiency and fake rates of less than
5\%. These systems are followed by the solenoid coil and then
a $K^0_L$ and muon detection system (KLM) that identifies
muons with less than 2\% fake rate
above 1 GeV/c. As a hadron absorber it also detects $K_L^0$
showers with an angular resolution of a few degrees.

The Belle detector has operated reliably over the first
two years of KEK-B operation, accumulating
78 fb$^{-1}$ on the $\Upsilon(4S)$
resonance corresponding to 85 million $B {\bar B}$
pairs that can be used to study $CP$ violation
in $B^0$ decay~\cite{phi1_prd}.

\section{The Measurement of $\sin 2 \phi_1$}

\begin{figure}
\begin{center}
\vspace*{-0.6in}
\mbox{\epsfxsize 12cm \epsfbox{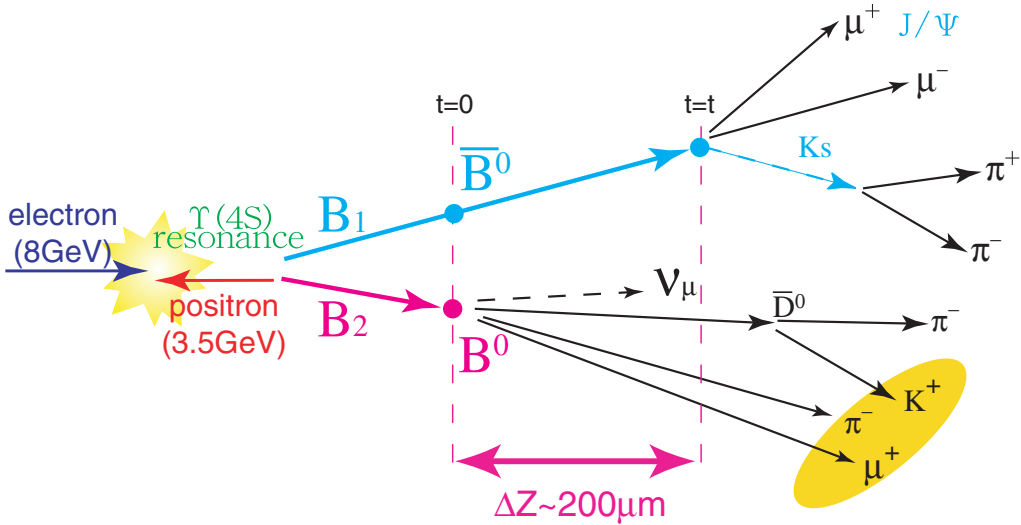}}
{\caption [CP violation cartoon ]
{\label{cartoon}
A schematic view the ingredients necessary to
observe $CP$ violation in $B$ meson decay at
an asymmetric $e^+e^-$ collider. }}
\end{center}
\end{figure}

There are three main ingredients that go into the measurement
of a $CP$ phase -- for example $\sin 2\phi_1$ in
$J/\psi K_S^0$ decay -- at Belle. These are illustrated in
fig.~\ref{cartoon}. First, one must identify the decay
vertices of the two $B$ mesons. Second, since
the final state under study is a $CP$ eigenstate,
we must tag the flavour of one
$B$ meson when it decays. Since the $\Upsilon(4S)$ decay
products evolve in a coherent $J=1$ state, the decay of
one of the $B$ mesons into a $B^0$ (in the example shown
in fig.~\ref{cartoon})
projects the other $B$ meson into a known ${\bar B^0}$ state.
Finally, one must identify a sample of candidates
which are $CP$ eigenstates -- for example $J/\psi K^0_S$ decays.
I will briefly describe each of these aspects of the
measurement in the following sections.

\subsection{$B$ Meson Decay Vertexing}
\label{sec:vertex}

Identifying and separating the two $B$ decay
vertices on an event-by-event basis in Belle is a subtle
process. Despite the boost from the asymmetric
beam energies the $B$ mesons only travel
a few hundred microns ($\gamma c \tau \approx 200 \mu$m) 
on average before they decay.
The silicon vertex detector in Belle 
measures the decay vertices with a precision of
about 100 $\mu$m. A $J/\psi K^0_S$
decay vertex is typically determined
with a precision of 75 $\mu$m, slightly better than
the flavour tag $B$ decay, which has a precision of 
140 $\mu$m, because there
are generally more reconstructed tracks attached to the
former vertex.

\begin{figure}
\begin{center}
\vspace*{-0.6in}
\mbox{\epsfxsize 7cm \epsfbox{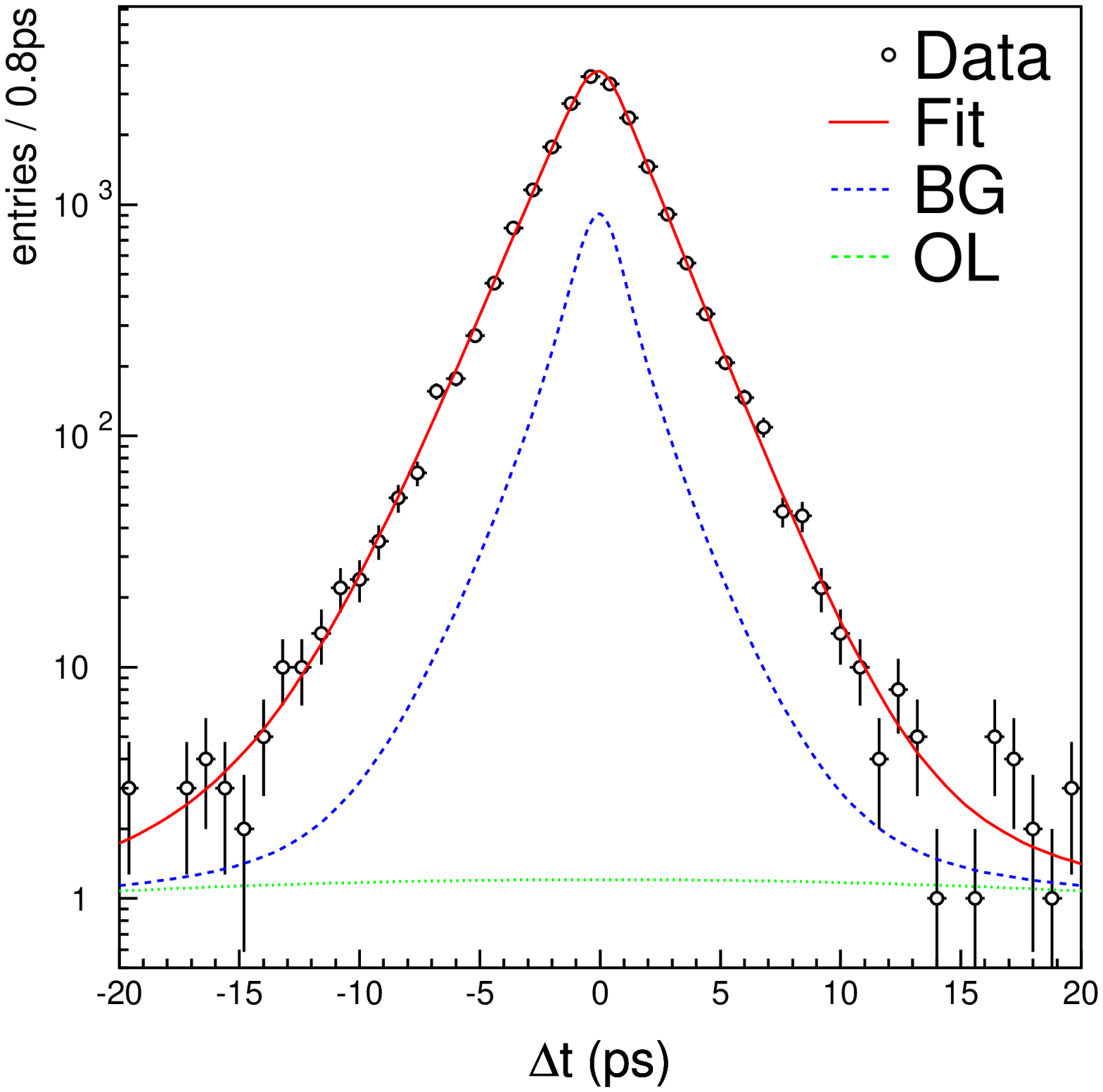}} 
\mbox{\epsfxsize 7cm \epsfbox{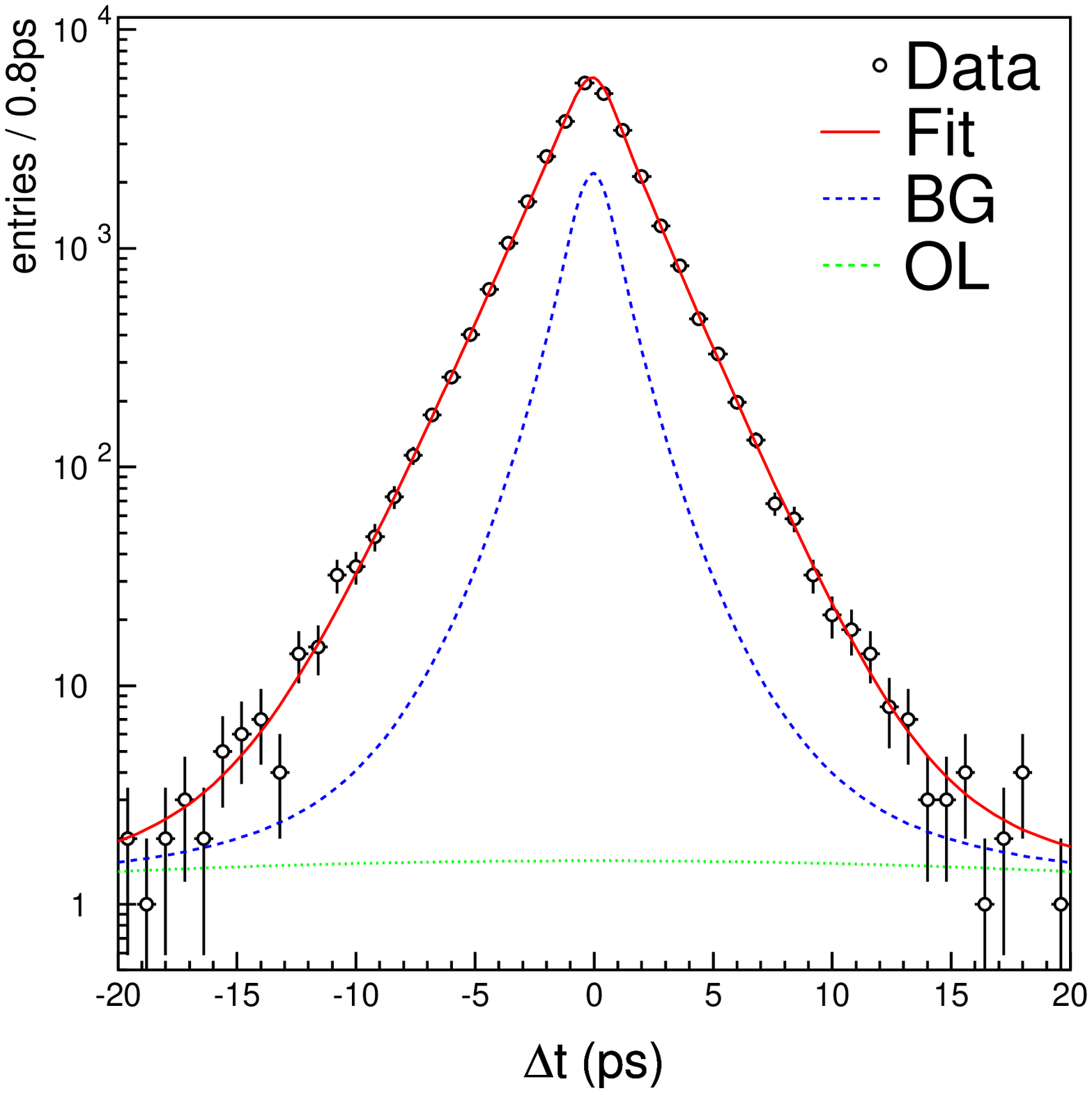}}
{\caption [The Belle lifetime measurements]
{\label{b_life}
Left: The distribution of decay time differences for $B^0$
and ${\bar B^0}$ mesons measured by Belle.
The open circles are the data, the solid line
is the result of a maximum likelihood fit that includes
a component proportional to the lifetime, and two components
(dashed and dotted) that parametrise the detector
resolution. 
Right: The distribution of decay time differences for $B^+$
and $B^-$ mesons.}}
\end{center}
\end{figure}   

The precision of Belle's vertexing is demonstrated by its
$B$ lifetime measurements (shown in fig.~\ref{b_life}) where
one sees a clear difference
between the detector resolution (dashed line) and the
longer lived $B$ meson decays (solid line through the
data points). These fits give precise measurements
of the charged and neutral $B$ meson lifetimes and their
ratio: $\tau_+ / \tau_0 = 1.09 \pm 0.03$~\cite{life_prl}.
The detector resolution is understood out to ten $B$ lifetimes
and over three orders of magnitude in $B$ decay rate. This is
one of the crucial ingredients to measuring $CP$
violation.

\subsection{Flavour Tagging in $B^0$ Meson Decay}
\label{sec:tag}

\begin{figure}
\begin{center}
\vspace*{-0.2in}
\mbox{\epsfxsize 10cm \epsfbox{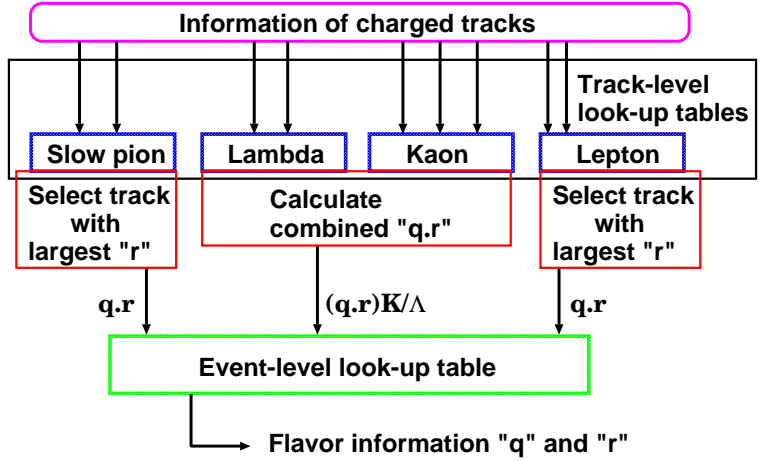}}
{\caption [The Belle flavour tagging algorithm]
{\label{ftag}
A schematic view of the Belle $B$ flavour tagging
algorithm. Events are classified by the
following characteristics: charged leptons, 
kaons, lambdas and slow pions. This
information is sifted through a set of look-up tables 
(based on Monte Carlo simulation) to identify and quantify the
most reliable flavour tagging information in each event. 
The result of this procedure is the assignment
of a ``charge'', $q$,
and the reliability, $r$, of the flavour assignment. }}
\end{center}
\end{figure}

A second key ingredient is being able to tag the flavour
of the $B$ meson accompanying the $CP$ eigenstate in the
$B {\bar B}$ decay. This is done without fully reconstructing
the opposite $B$ -- which would result in a significant
loss in efficiency. Instead we look for characteristics of
the other $B$ decay that tag its flavour. For example,
high momentum leptons indicate a direct semi-leptonic
$B$ decay, where the charge of the lepton is determined by
the flavour of $B$. Lower momentum leptons arise
from cascade semi-leptonic decays where the $B$
meson first decays to a charmed meson and the latter decays
semi-leptonically. This results in the opposite correlation
between the lepton charge and $B$ meson flavour. Such
information is combined in a set of look-up tables shown
in fig.~\ref{ftag} that classify potential
flavour information according to the sign of the $b$
quark charge, $q$, and the reliability of the tagging
information, $r$.

\begin{figure}
\begin{center}
\vspace*{-0.6in}
\mbox{\epsfxsize 12cm \epsfbox{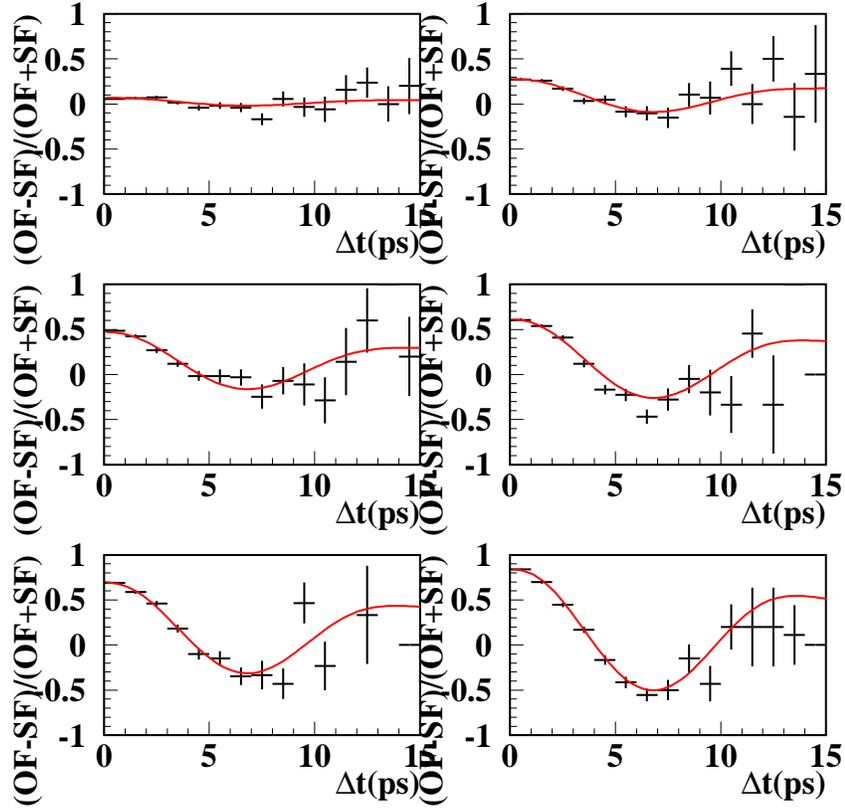}}
{\caption [B mixing measurements at Belle]
{\label{bmix}
$B$ mixing measurements in self-tagged $B^0 \rightarrow D^{*+} l^- \nu$
samples. This sample calibrates our 
tagging efficiency. The amplitude of the mixing 
observed, hence the efficiency of our tagger, 
depends on the reliability of the tag,
which increases from top-left to bottom-right in
this series of plots. }}
\end{center}
\end{figure}

While the look-up table information: $q$, $r$ and their
correlations -- when more than one piece of tagging information
is available in a single event -- is extracted from a
Monte Carlo simulation, the efficiency of the 
tagging algorithm is calibrated
on a data control sample. Using a sample of $B \rightarrow D^* l \nu$
decays -- where one knows the flavour of the $B$ meson
from the charge of the lepton in the final state
-- the $B$ mixing
parameter is measured from the time-dependence of the observed decays. 
We use our flavour tagging algorithm, described above, to
classify events into six different ranges of $r$, shown in fig.~\ref{bmix}.
One sees that the observed amplitude of the mixing oscillation is
much less in the sample tagged with lowest reliability
(top left) while the amplitude of the oscillation (at $\Delta t = 0$)
almost reaches 1 for those events that our algorithm tags with
the highest reliability (bottom right).

\begin{figure}
\begin{center}
\vspace*{-0.6in}
\mbox{\epsfxsize 10cm \epsfbox{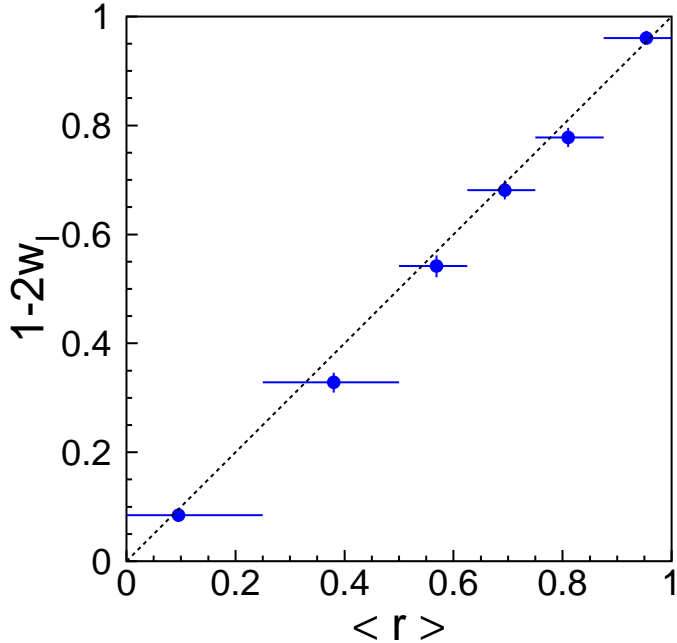}}
\vspace*{-0.3in}
{\caption [Tagging Calibration ]
{\label{tag_cal}
The correct tag probability ($1 - 2 w_l$) versus
tag reliability, $r$. The wrong-tag fraction ($w_l$)
is measured from the observed mixing 
in self-tagged $B^0$ samples (see fig.~\ref{bmix}).
The measured tagging probabilities ($1 - 2 w_l$)
are used to weight the flavour determination in
fits to $CP$ eigenstates.}}
\end{center}
\end{figure}

Figure~\ref{tag_cal} shows the correct tag probability from the six
plots in fig.~\ref{bmix} versus the average $r$ for each sample.
The correct tag probability, $1 - 2 w_l$, is extracted from 
$w_l$ -- the probability that a tag gives the wrong charge sign.
In cases were $w_l$ approaches 0.5 (a 50-50 guess at the charge
of the $b$ quark) the probability of correctly tagging the flavour of the $B$
meson goes to 0. Conversely, when $w_l$ goes to
0, one approaches a 100\% correct tag probability. Figure~\ref{tag_cal}
shows a strong correlation between our tagger's reliability, $r$,
and the correct tag probability as measured in our control sample.
These measured tag probabilities are used to weight the $CP$ eigenstate 
decays in the fit to extract their decay asymmetry.

With this algorithm we extract some tagging information from 99.5\%
of all $B$ decay candidates in Belle. While some of the events have tags with low
reliability, we measure an overall tagging efficiency of
$28.8 \pm 0.6 \%$, corresponding to 
almost 1/3 of our $B$ sample being perfectly tagged. This represents
an improvement from an effective tagging efficiency of $27.7 \pm 1.2 \%$ 
in previous measurements~\cite{phi1_prd} arising from an increase 
to our low-momentum track reconstruction efficiency and an improved
silicon alignment -- allowing us to associate more
tracks with the tagging $B$ decay vertex. The reduced uncertainty on
our tagging efficiency is the result of 
tripling of the size of the control data sample.
 
\subsection{$CP$ Eigenstate Event Samples}

\begin{figure}[ht]
\begin{center}
\vspace*{-0.6in}
\mbox{\epsfxsize 10cm \epsfbox{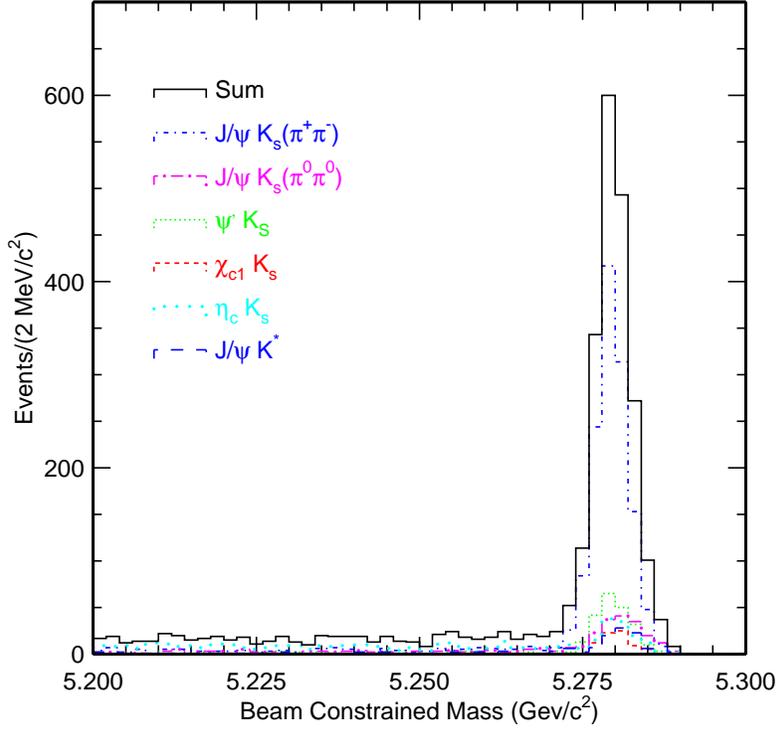}}
\vspace*{-.8in}
{\caption [Psi K_S modes ]
{\label{ccbar_short}
The beam constrained mass for the different
$B^0 \rightarrow {\bar c}c K_S^0$ final states used to measure indirect
$CP$ violation in Belle. The largest
sample comes from the $J/\psi K^0_S (\pi^+\pi^-)$ decays
(dot-dashed line). The other modes combine to make up
about 20\% of our sample (see table~\ref{cp_candidates}).}}
\end{center}
\end{figure}

From 78 fb$^{-1}$ of data taken at the $\Upsilon(4S)$
resonance Belle has identified
a sample of almost 3000 $B^0 \rightarrow c{\bar c} K$ decay candidates. These
are summarised in table~\ref{cp_candidates} where one sees that
a little over 1/3 of the candidates come from $J/\psi K^0_S$ decays
where the $K^0_S$ is reconstructed in a $\pi^+\pi^-$ final
state. This golden mode also has the highest purity. We also have a 
number of other modes that are used
to measure $\sin 2 \phi_1$ with other $c {\bar c}$
resonances and other $K^0_S$ final states. Finally, 
about $40 \%$ of our sample is in the form of $J/\psi K_L^0$
decays that have the opposite $CP$ eigenvalue. This sample
provides an important cross-check for possible $CP$ dependent
systematic effects in 
our measurement. Figure~\ref{ccbar_short} shows the mass
distribution for our candidates with $K^0_S$ candidates in
the final state. 

\renewcommand{\arraystretch}{1.1}
\begin{table}
\begin{center}
\begin{tabular}{l|ccc}
\hline \hline
Mode                        & CP ($\xi_K$) & Candidates & Purity (\%) \\ \hline
$J/\psi K_S^0 (\pi^+\pi^-)$ &  -1          &   1116     &   98        \\
$J/\psi K_S^0 (\pi^0\pi^0)$ &  -1          &    162     &   82        \\
$\psi(2S) K_S^0$            &  -1          &    172     &   93        \\
$\chi_{c1} K_S^0$           &  -1          &     67     &   96        \\
$\eta_c K_S^0$              &  -1          &    122     &   68        \\ 
$J/\psi K^{*0} (K_S^0\pi^0)$&  1 (81\%)    &  89     &   92        \\ 
$J/\psi K_L^0$              &  1           &   1230     &   63        \\ \hline \hline
 Total                      &              &   2958     &             \\ \hline \hline
\end{tabular}
\end{center}
\caption{Summary of the $CP$ candidates, after flavour tagging and decay
vertexing, used to measure $\sin 2 \phi_1$.}
\label{cp_candidates}
\end{table}

\begin{figure}
\begin{center}
\vspace*{-0.6in}
\mbox{\epsfxsize 10cm \epsfbox{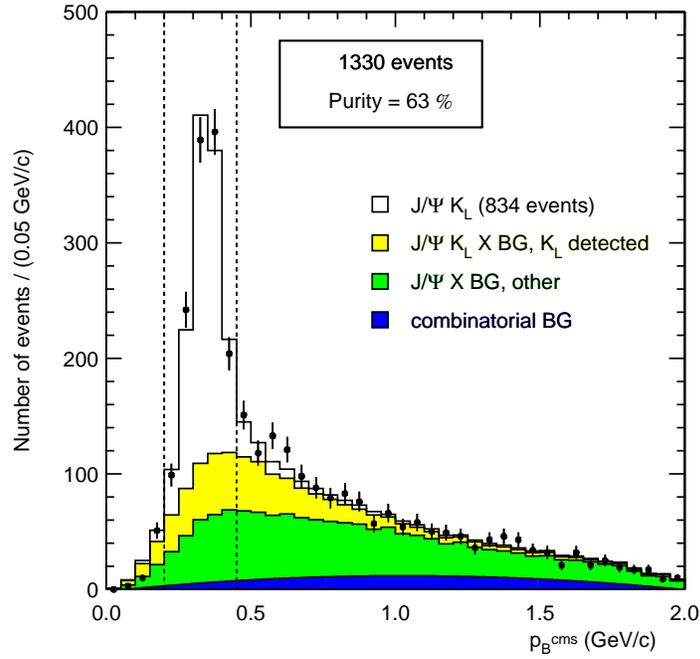}}
\vspace*{-0.2in}
{\caption [Psi K_L modes ]
{\label{ccbar_long}
The centre of mass momentum distribution for Belle's
$J/\psi K_L^0$ candidates used to measure indirect
$CP$ violation. The signal
forms a peak at $p_B^{\rm cms} = 0.33$ GeV/c, while
backgrounds -- where one or more final state particles
have not been included in the $B^0$ candidate
reconstruction -- have their momentum smeared out.}}
\end{center}
\end{figure}

While our $K_L^0$ detector does not measure the energy of
the hadronic shower of a neutral Kaon decay with high
precision, it does measure the shower direction 
with an accuracy of a few degrees. Having
fully reconstructed a $J/\psi$ meson one can hypothesize
that a $B^0 \rightarrow J/\psi K_L^0$ decay has occurred
and infer the magnitude of the $K_L^0$ momentum using
only the $K_L^0$ direction, by constraining the
two body system ($J/\psi$, $K_L^0$) to have the $B^0$
mass. Imposing this constraint reduces the number of handles one has to
reject background but still provides one degree
of freedom -- chosen to be the momentum of the 
$B^0$ candidate in $\Upsilon(4S)$ centre of momentum
(shown in fig.~\ref{ccbar_long}). Successfully reconstructed $J/\psi K_L^0$
candidates peak near $p_B^{\rm cms} \approx 0.3$ GeV/c,
while backgrounds generate a flatter distribution
in $p_B^{\rm cms}$. From the distribution in fig.~\ref{ccbar_long}
we extract 1330 candidate events (before flavour tagging or
decay vertex reconstruction) and estimate their
purity to be 63\%. Since many of the backgrounds under
the $J/\psi K_L^0$ peak result from other $B$ meson decays
care must be taken when extracting a $CP$ asymmetry from this
sample to account for the $CP$ asymmetry of the background.
We do this with Monte Carlo studies whose ability to constrain
the $CP$ content of the background is reflected in a systematic 
uncertainty on $\sin 2 \phi_1$ from this channel.

\subsection{Results}

We perform a maximum likelihood fit to all our $B^0 \rightarrow c {\bar c}
K$ candidates to extract their time dependent asymmetry
(as shown in eqn.~\ref{asym}). This fit includes the
flavour tagging probability, uncertainties on the
measured separation between the two $B$ meson decays
(converted to $\Delta t$ using the $B$ meson boost)
as well as the purity of the signal for each type 
of decay. The $J/\psi K_S^0 (\pi^+\pi^-)$ candidates receive
the highest weight because they have much smaller backgrounds
(see table~\ref{cp_candidates}).

\begin{figure}
\begin{center}
\vspace*{-0.5in}
\mbox{\epsfxsize 10cm \epsfbox{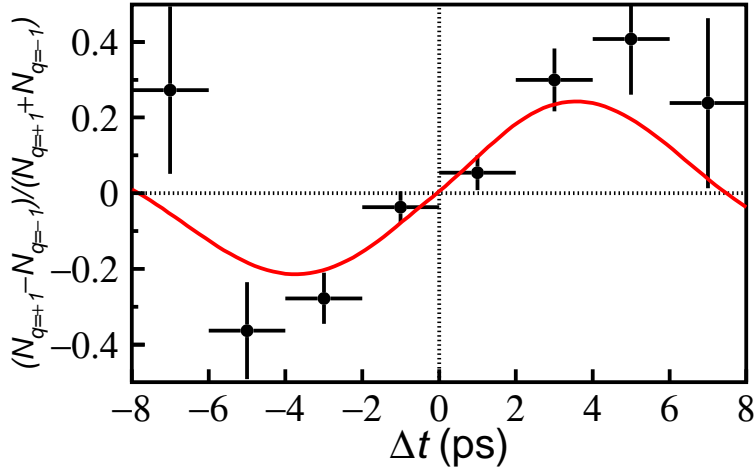}}
{\caption [CP fit - K_S]
{\label{cpminus}
The time dependent asymmetry for $B^0$ and 
${\bar B^0}$ decays to ${\bar c}c K_S^0$
($\xi_{\rm CP} = -1$). The amplitude of this
oscillation, extracted from an event-by-event
maximum likelihood fit, gives 
$\sin 2\phi_1 = 0.716 \pm 0.083$.}}
\end{center}
\end{figure} 

\begin{figure}
\begin{center}
\vspace*{-0.8in}
\mbox{\epsfxsize 10cm \epsfbox{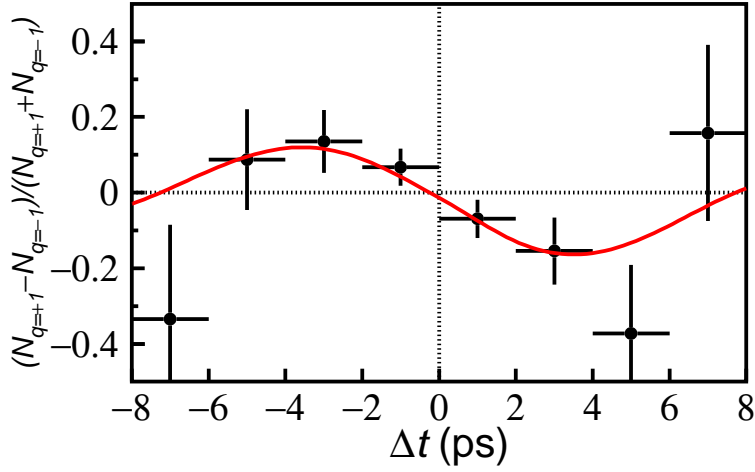}}
{\caption [CP fit - K_L]
{\label{cpplus}
The time dependent asymmetry for $B^0$ and 
${\bar B^0}$ decays to $J/\psi K_L^0$ 
($\xi_{\rm CP} = 1$). The amplitude of this
oscillation, extracted from an event-by-event
maximum likelihood fit, gives 
$\sin 2\phi_1 = 0.781 \pm 0.167$.}}
\end{center}
\end{figure} 

\begin{figure}
\begin{center}
\vspace*{-0.6in}
\mbox{\epsfxsize 10cm \epsfbox{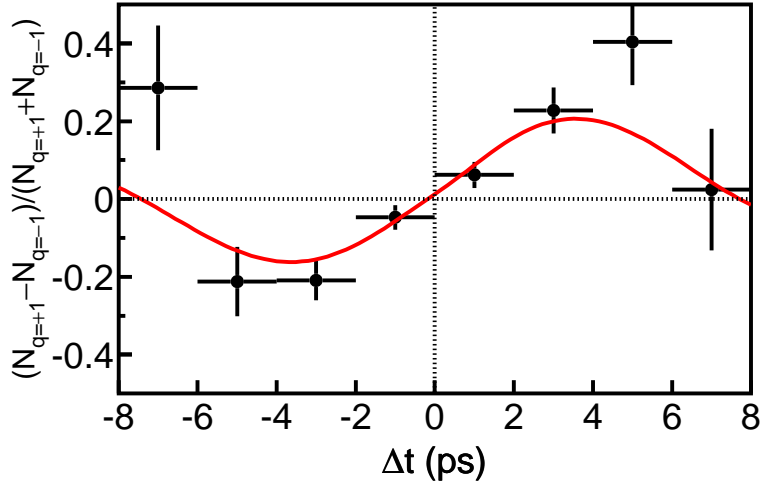}}
{\caption [CP fit - all]
{\label{cpall}
The combined time dependent asymmetry for $B^0$ and 
${\bar B^0}$ decays to all ${\bar c} c  K^0$ final
states (flipping the sign of the candidates with
$\xi_{\rm CP} = 1$). An event-by-event
maximum likelihood fit gives 
$\sin 2\phi_1 = 0.719 \pm 0.074$.}}
\end{center}
\end{figure}

We first perform this fit separately for candidates with
$CP = -1$ (the bulk of our data) and $CP = +1$ (our
$J/\psi K^0_L$ sample), shown in figs.~\ref{cpminus} 
and~\ref{cpplus}, respectively. We find the magnitude
of the fit asymmetries are equal, but they have opposite
sign -- as expected. The fit results are reported
in table~\ref{cross_check}. We then combine all
candidates into a single fit (see fig.~\ref{cpall}) 
-- inverting the sign of $\Delta t$ for the $CP = +1$ 
candidates -- and obtain:
$$ \sin 2 \phi_1 = 0.719 \pm 0.074 (stat.) \pm 0.035 (sys.)$$

The systematic uncertainties involved in this measurement
are listed in table~\ref{cp_sys}. The largest of
these uncertainties
are derived from measurements of our control
samples -- for example the flavour tagging discussed in 
section~\ref{sec:tag} -- and thus can be expected to
shrink as more data becomes available~\cite{phi1_ichep_prl}.
The overall systematic
uncertainty is small compared to the statistical
precision which bodes well for future measurements.

\renewcommand{\arraystretch}{1.1}
\begin{table}[ht]
\begin{center}
\begin{tabular}{cc}
\hline \hline
Uncertainty Source         & Value    \\ 
\hline
 Vertexing Reconstruction  &  0.022   \\
 Flavour Tagging           &  0.015   \\
 Vertex Resolution         &  0.014   \\
 Fit parametrisation       &  0.011   \\ 
 $J/\psi K_L^0$ Background &  0.010          \\
 $\Delta m_d$, $\tau_B$    &  $\leq 0.010$   \\ 
\hline \hline
 Total                     &  0.035   \\ 
\hline \hline
\end{tabular}
\caption{Systematic uncertainties on the measurement of $\sin 2 \phi_1$.}
\label{cp_sys}
\end{center}
\end{table}

\subsection{ Cross-checks of $\sin 2 \phi_1$ }

We perform a number of cross-checks on non-$CP$ eigenstate
samples, all of which exhibit null asymmetries with 
statistical precisions ranging from $0.02$ to $0.09$. We do
not include these as systematic uncertainties as we find
no evidence of bias but they
provide additional confidence in our main result.
We have also sub-divided our data into different sub-samples
to see whether there is any evidence for systematic
variations in our result. Finding none (see table~\ref{cross_check})
gives us further confidence in our main result but again
do not
ascribe additional systematic uncertainty as  the
variations are all consistent with our quoted
result within the 
statistics of the smaller sub-samples involved.

\renewcommand{\arraystretch}{1.1}
\begin{table}[ht]
\begin{center}
\begin{tabular}{lc}
\hline \hline
Subsample (stat error only)                                       &                     \\
\hline    
  $J/\psi K_S^0 (\pi^+\pi^-)$                                     &  $ 0.73 \pm 0.10$   \\ 
  $(c{\bar c}) K_S^0$ (except  $J/\psi K_S^0 (\pi^+\pi^-)$)       &  $ 0.67 \pm 0.17$   \\ 
  $J/\psi K^0_L$                                                  &  $ 0.78 \pm 0.17$   \\ \hline
 $f_{tag} = B^0$                                                  &  $ 0.65 \pm 0.12$   \\ 
 $f_{tag} = {\bar B^0}$                                           &  $ 0.77 \pm 0.09$   \\ \hline
 $r \leq 0.5 $                                                    &  $ 1.26 \pm 0.36$   \\ 
 $ 0.5 \leq r \leq 0.75 $                                         &  $ 0.62 \pm 0.15$   \\ 
 $ 0.75 \leq r $                                                  &  $ 0.72 \pm 0.09$   \\  \hline \hline 
 All                                                              &  $ 0.72 \pm 0.07$   \\ \hline \hline
\end{tabular}
\caption{Cross-checks of $\sin 2 \phi_1$ from fits to subsets of the data.}
\label{cross_check}
\end{center}
\end{table}

\section{$CP$ Asymmetries in $B^0 \rightarrow \pi^+\pi^-$ Decay}

\begin{figure}
\begin{center}
\vspace*{-0.6in}
\mbox{\epsfxsize 7cm \epsfbox{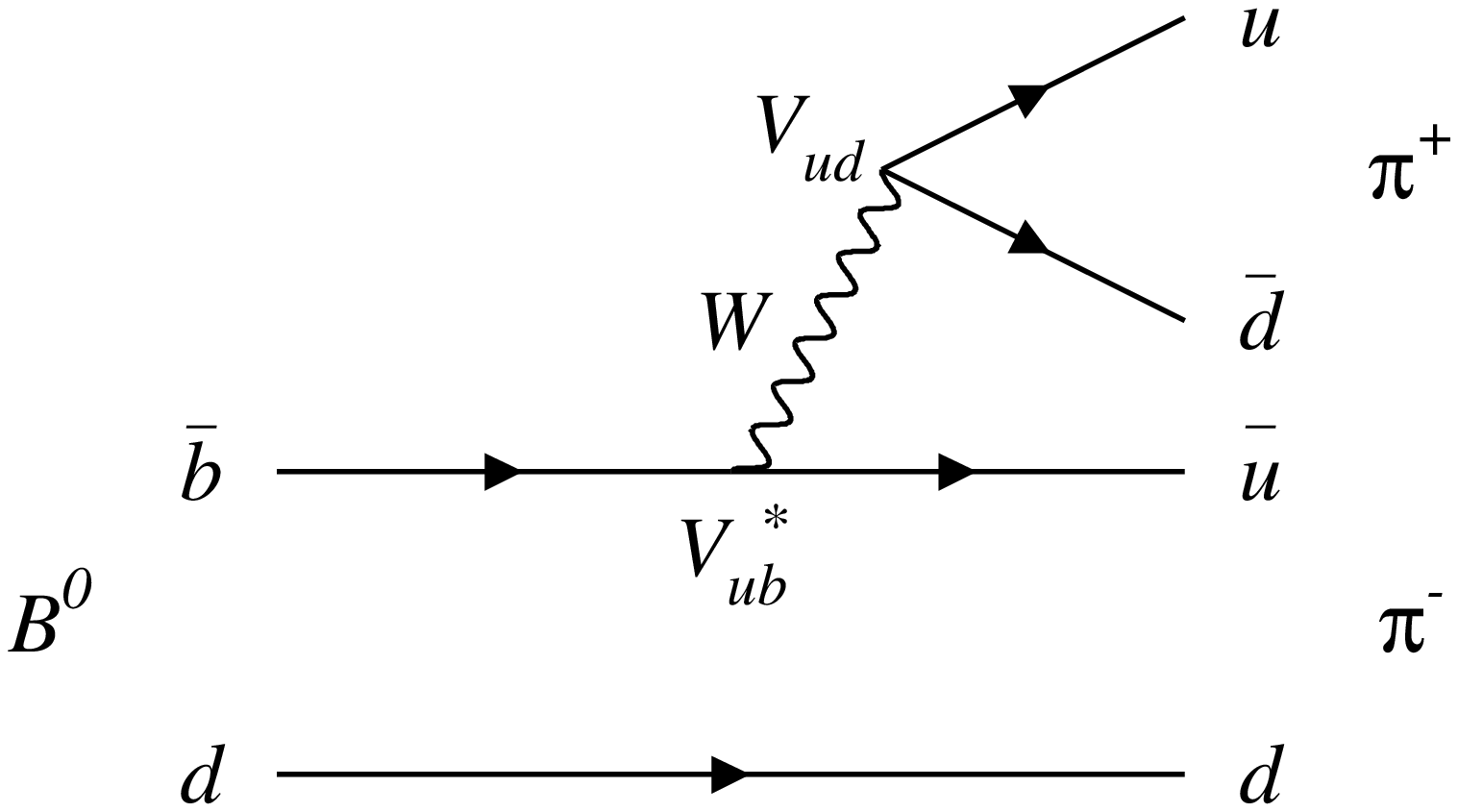}}
\mbox{\epsfxsize 7cm \epsfbox{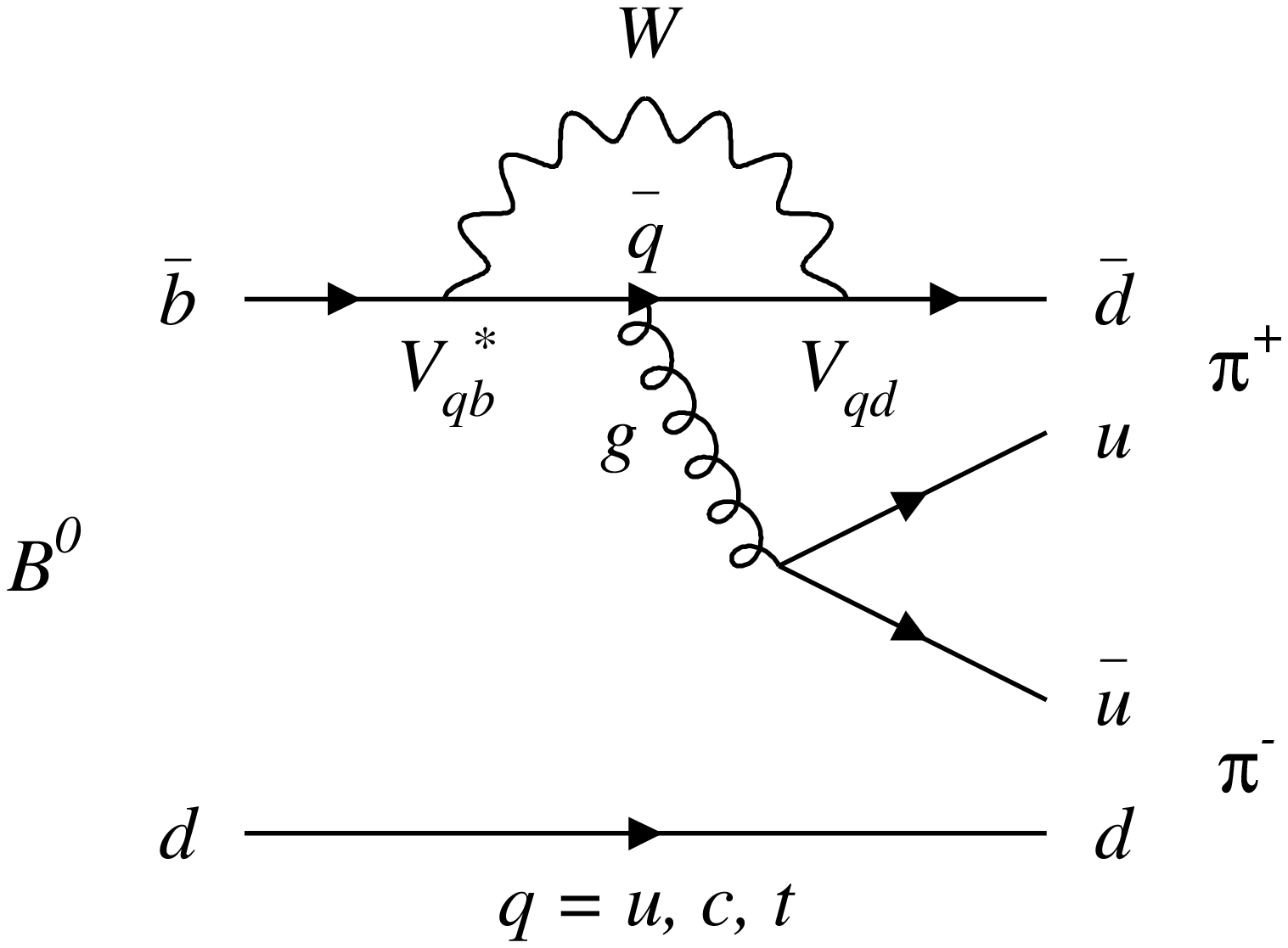}}
\vspace*{-0.6in}
{\caption [Pi Pi final state diagrams]
{\label{pipi_diagrams}
Two tree level diagrams for $B^0 \rightarrow \pi^+ \pi^-$
decays. Given that there are two paths to the same
$CP$ eigenstate, their interference can produce direct
$CP$ violation in this mode. }}
\end{center}
\end{figure}

Having observed indirect $CP$ violation in $B \rightarrow J/\psi K$
decays it is natural to study other 
$CP$ eigenstates. The decay $B^0 \rightarrow \pi^+\pi^-$ is 
interesting because there are at least two significant amplitudes
that can interfere in the direct decay. These amplitudes are show diagrammatically
in fig.~\ref{pipi_diagrams}. With more than one amplitude
in the direct decay path one can have a $\cos \Delta m \Delta t$
dependence in the $CP$ asymmetry. This dependence can come from 
the interference of the direct
amplitudes and can appear in addition to the $\sin \Delta m \Delta t$
time-dependence (compare for example to eqn.~\ref{asym}) 
from the interference with the mixing amplitude:
\begin{center}
\begin{tabular}{ccl}
$A_{CP}(\Delta t)$ & = & ${ { {dN \over dt} (\bar{B^0} \rightarrow F_{\rm CP}}) - 
               {dN \over dt} (B^0 \rightarrow F_{\rm CP}) } \over 
               { {dN \over dt} (\bar{B^0} \rightarrow F_{\rm CP}) + 
               {dN \over dt} (B^0 \rightarrow F_{\rm CP}) }$, \\ 
            & = & $S_{F} \sin~\Delta m \Delta t + A_{F} \cos~\Delta m \Delta t. \qquad \qquad [2]$
\label{asym_2}
\end{tabular}
\end{center}
The $B^0 \rightarrow K^+ \pi^-$ branching
fraction -- three times larger than the $B^0 \rightarrow \pi^+ \pi^-$
branching fraction -- is strong evidence that the penguin
amplitude (right diagram in fig.~\ref{pipi_diagrams}) is
not small. In the decay $B^0 \rightarrow \pi^+ \pi^-$ the CKM 
elements involved predict an asymmetry proportional
to the angle $\phi_2$. However the 
penguin diagram can introduce a non-CKM phase thus:
$$ S_F \approx \sin(2 \phi_2 + \theta) $$
where $\theta$ can come from phases at the gluon vertices 
in the penguin process.
While waiting for measurements of the various $B \rightarrow \pi \pi$
branching fractions~\cite{brendan} with sufficient precision
to constrain $\theta$, it is still interesting to measure
the $\pi^+\pi^-$ asymmetry and perhaps get some hint for the
size of the combination $\sin(2 \phi_2 + \theta)$~\cite{Rosner}.

\subsection{$B^0 \rightarrow \pi^+\pi^-$ Candidate Selection}

\begin{figure}[ht]
\begin{center}
\vspace*{-0.6in}
\mbox{\epsfxsize 10cm \epsfbox{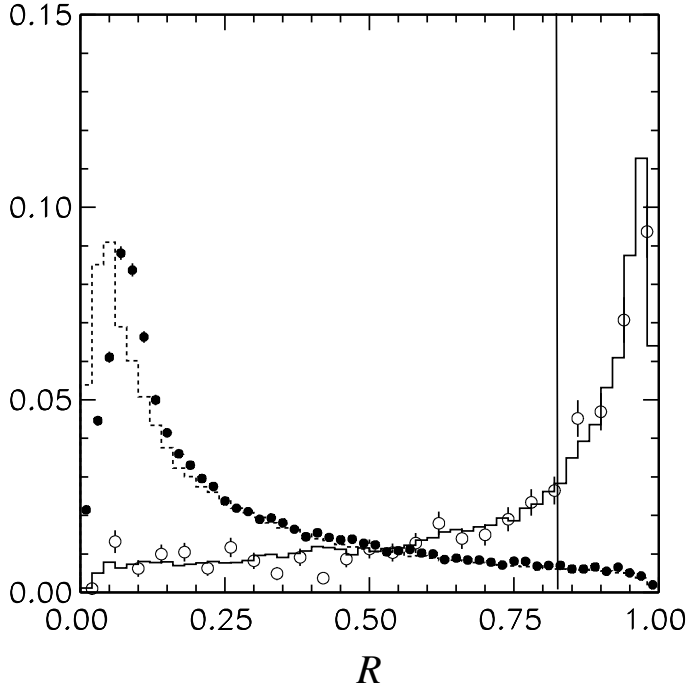}}
\vspace*{-0.2in}
{\caption [Continuum suppression]
{\label{cont_supp}
Low multiplicity final states are subject to higher
background from the $e^+e^-$
continuum. To
suppress these Belle combines event shape variables
to get an overall likelihood, $R$. The
solid points are continuum data, while the open
points are a control sample of fully reconstructed,
low-multiplicity, $B$ candidates. $B^0 \rightarrow \pi^+\pi^-$
candidates are required to have $R > 0.8$. }}
\end{center}
\end{figure}

The candidates and result presented here are from 
43 fb$^{-1}$ of data collected by Belle~\cite{pipi_prl}.
While the $B$ vertexing and flavour tagging described in
sections~\ref{sec:vertex} and~\ref{sec:tag} can be used in 
a measurement of the $\pi^+\pi^-$ 
asymmetry, the candidate selection is
more challenging. The branching fractions for
$B \rightarrow h^+h^-$ are an order of magnitude
smaller than those of the $c {\bar c} K$ final states -- on the
order of $10^{-5}$. Furthermore the $B \rightarrow h^+ h^-$
decays have only two, relatively high momentum, tracks in the
final state. Such candidates are much more prone to look like the
$e^+e^- \rightarrow q {\bar q}$ continuum background.
Belle uses a set of event-shape variables (Fox-Wolfram moments)
and the reconstructed $B^0$ candidate direction to 
suppress continuum background. These are combined into a
single variable, $R$, shown in figure~\ref{cont_supp} 
for $B {\bar B}$ data (open circles) and off-resonance
data (closed circles). By requiring $R > 0.8$ we reduce
the continuum background by an order of magnitude while 
retaining two thirds of our $B$ candidates.

\begin{figure}[ht]
\begin{center}
\vspace*{0in}
\mbox{\epsfxsize 6.5cm \epsfbox{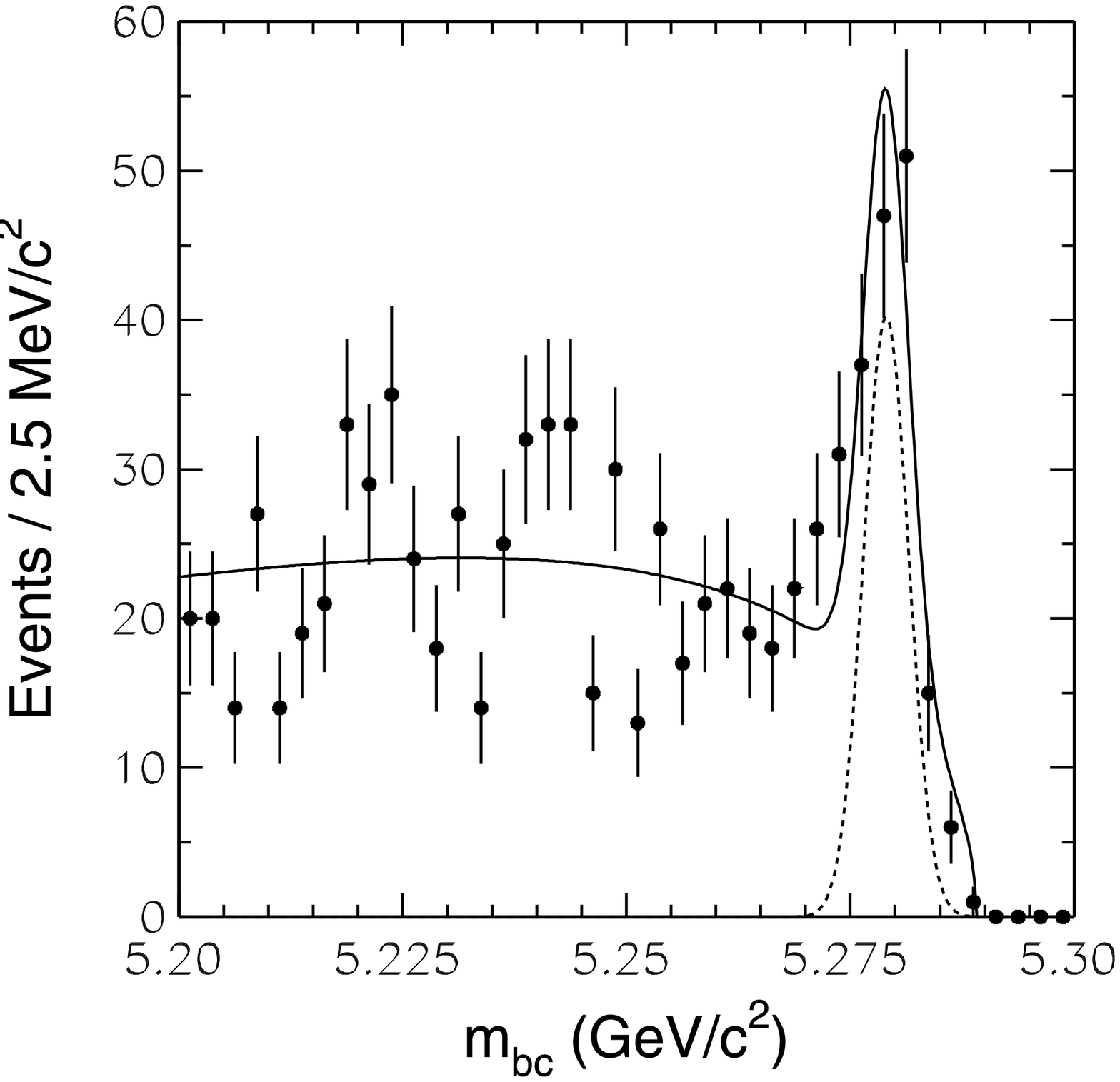}}
\mbox{\epsfxsize 6.2cm \epsfbox{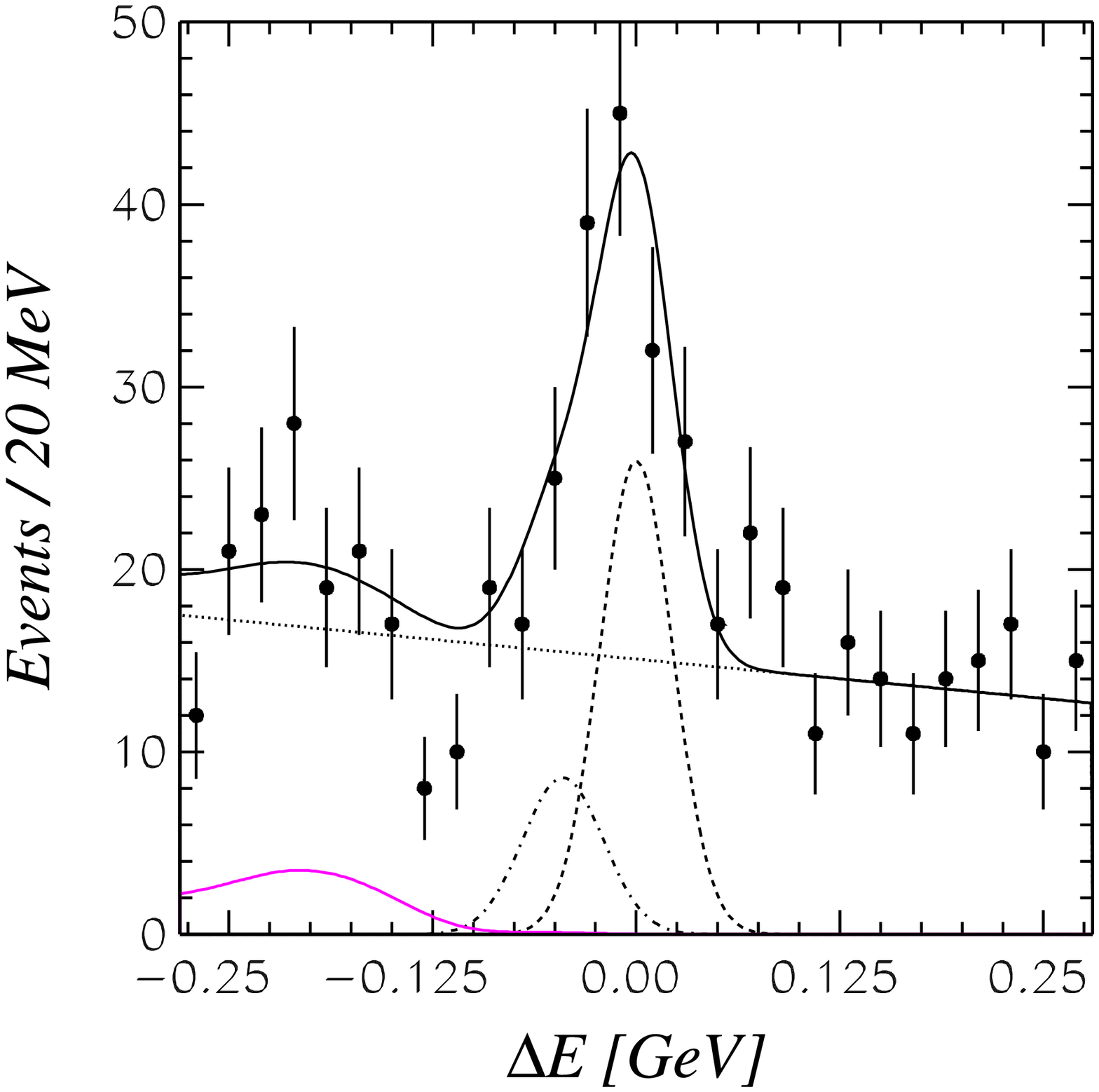}}
{\caption [Pi Pi candidates, beam constrained mass]
{\label{B to hh mass}
Left: The beam constrained mass, m$_{\rm bc}$, for the $B^0 \rightarrow \pi^+\pi^-$ candidates
used to extract the $CP$ asymmetry. Right: The energy difference distribution, $\Delta E$, 
for the same candidates. In $\Delta E$ the $\pi^+\pi^-$ signal
peaks at 0 (dashed Gaussian), the $K^+ \pi^-$ background peaks
at -0.045 GeV (dot-dashed Gaussian).
The continuum is parametrised as a falling linear background
(dotted line) while $B^0$ decays with more that two particles in the
final state appear at low $\Delta E$.}}
\end{center}
\end{figure}

Since these events are fully reconstructed we are able
to measure both the beam-constrained mass, m$_{\rm bc}$,
and the difference between the reconstructed energy
of the $B$ candidate and the beam energy, $\Delta E$. Figure~\ref{B to hh mass}
shows the distributions of both of these 
variables for our $B \rightarrow h^+ h^-$ sample.
Comparing the m$_{\rm bc}$ distribution for these candidates
to that of the $c {\bar c} K$ candidates in fig.~\ref{ccbar_short}
one sees that the background is much
more prominent. In the $\Delta E$ distribution (on the
right in fig.~\ref{B to hh mass}) one can see the source of 
these backgrounds. The remaining continuum
background populates $\Delta E$ almost uniformly --
falling linearly with increasing $\Delta E$ due to phase
space -- and is represented by the dotted line. 
$B$ decays to three or more particles,
where we have missed one in our reconstruction,
populate the
low $\Delta E$ region (grey line) at a much lower
rate than the continuum. Finally, $B^0 \rightarrow
K^+ \pi^-$ candidates that have not been rejected
by our particle identification produce a
peak at $\Delta E \approx -0.045$ GeV. This peak
is shifted because these candidates have been reconstructed
assuming that both $B^0$ decay daughter tracks 
are pions (with a mass of 139 MeV/c$^2$). If
one of them is really a kaon (with a mass of 494 MeV/c$^2$)
then the reconstructed energy shifts down by about 45 MeV.
This background is shown by the dot-dashed line.

We restrict our $CP$ fit to candidates that have $|\Delta E| < 0.067$ GeV,
eliminating the mis-reconstructed $B$ background entirely
and leaving us with 74 $\pi^+\pi^-$ candidates, 28 $K^+ \pi^-$
candidates and about 100 candidates that come from
the $e^+e^-$ continuum.

\subsection{The $CP$ Fit}

\begin{figure}[ht]
\begin{center}
\vspace*{1.0in}
\mbox{\epsfxsize 6.5cm \epsfbox{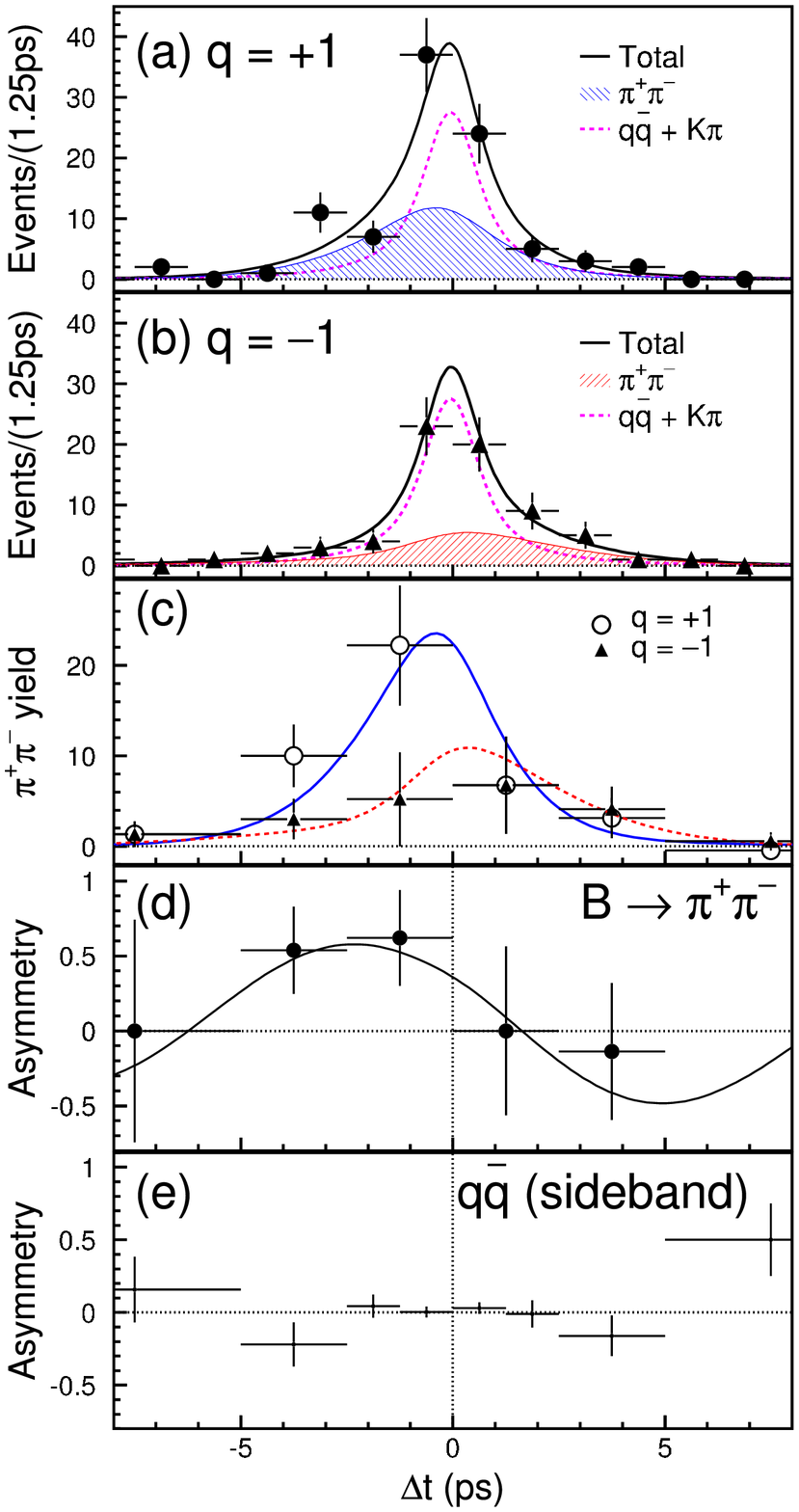}}
\vspace*{-0.3in}
{\caption [Pi Pi CP fit]
{\label{pipi_cpfit}
The results of the $\pi^+ \pi^- \> \> CP$ asymmetry fit. a) The $\Delta t$
distribution for candidates tagged with $q=1$. The signal is
shown as the hatched region, while the background (dominated
by the continuum) is shown by the dashed line. The solid
line is a projection of the overall fit and the points
are the data. b) The $\Delta t$ distribution for candidates
with tagged $q=-1$. c) The circles (triangles) are the 
background subtracted data for $q = (-)1$ while the curves
are fit projections for the two flavours. d) The difference
between the two data-samples in c). e) The
asymmetry observed in a continuum sample.}}
\end{center}
\end{figure}

We perform an event-by-event likelihood fit to the
time dependent asymmetry for the $B \rightarrow \pi^+ \pi^-$
candidates. In order to maximise our sensitivity to the
$\pi^+\pi^-$ asymmetry this fit weights the events according
to their continuum likelihood (events near $R=1$ 
are weighed more heavily) as well as their flavour
tagging probability and vertex reconstruction quality --
as was done for $\sin 2 \phi_1$.

Figures~\ref{pipi_cpfit} a) and b) show the $\Delta t$
distributions for events tagged as coming from $B^0$ and
${\bar B^0}$ decays, respectively. The continuum and $K^+\pi^-$
backgrounds are represented by the dashed lines, symmetric about
$\Delta t =0 $ in each case. The hatched areas in each plot
are what the fit ascribes to the $B \rightarrow
\pi^+\pi^-$ candidates. The sum of these two distributions give
the solid line which fits the data. 
Figure~\ref{pipi_cpfit} c) shows the $\Delta t$
distribution for just the $B \rightarrow \pi^+\pi^-$ component of the fit. 
Here we see that our fit predicts that there are
almost twice as many $B^0$ as ${\bar B^0}$ candidates 
in our sample -- independent of $\Delta t$ -- 
an indication of direct $CP$ violation.
Finally, fig.~\ref{pipi_cpfit} d) shows the asymmetry between
the two distributions in fig.~\ref{pipi_cpfit} c). Here one
also sees a $\Delta t$ dependent asymmetry leading to our fit
result:
\begin{center}
\begin{tabular} {c@{\extracolsep{2pt}}r@{}l@{}l}
$S_{\pi \pi}$ = & $-$ &$1.21 ^{+0.38} _{-0.27} ({\rm stat})$ & $^{+0.16} _{-0.13} ({\rm sys})$, \\
$A_{\pi \pi}$ = &  &$0.94 ^{+0.25} _{-0.31} ({\rm stat})$ & $\pm 0.09 ({\rm sys})$.  \\
\end{tabular}
\end{center}
This is three sigma evidence for both direct $CP$ violation in the
decay $B \rightarrow \pi^+\pi^-$ and indirect $CP$ violation~\cite{pipi_prl}.

We have performed a large number of cross-checks on this fit, including
fits to the $B^0 \rightarrow K^+\pi^-$ sample which should not have
a $CP$ violating asymmetry and fits to continuum side-bands 
(fig.~\ref{pipi_cpfit} e). None of these fits show
significant asymmetries. We have simulated the 
size of our statistical uncertainties. We find that our statistical 
uncertainty on $S_{\pi\pi}$ is somewhat smaller than might be
expected for a sample this size. Still there is a 5\% chance that
we could have gotten a smaller uncertainty.
While unlikely, this is not an unacceptable statistical fluctuation.
Finally, we have performed a series of toy Monte Carlo simulations,
each generated with $(S_{\pi\pi}, A_{\pi\pi}) = (0,0)$ and sample sizes
(signal and background) that mimic those in our data. We
find that 1.6\% of these toy Monte Carlo samples return a fit
value farther from $(S_{\pi\pi}, A_{\pi\pi}) = (0,0)$ than
our data. It will be interesting to see how this result evolves
as more data becomes available.

\section{$CP$ Asymmetries in Rarer $B^0$ Decay Modes}

\begin{figure}[ht]
\begin{center}
\vspace*{-0.6in}
\mbox{\epsfxsize 12cm \epsfbox{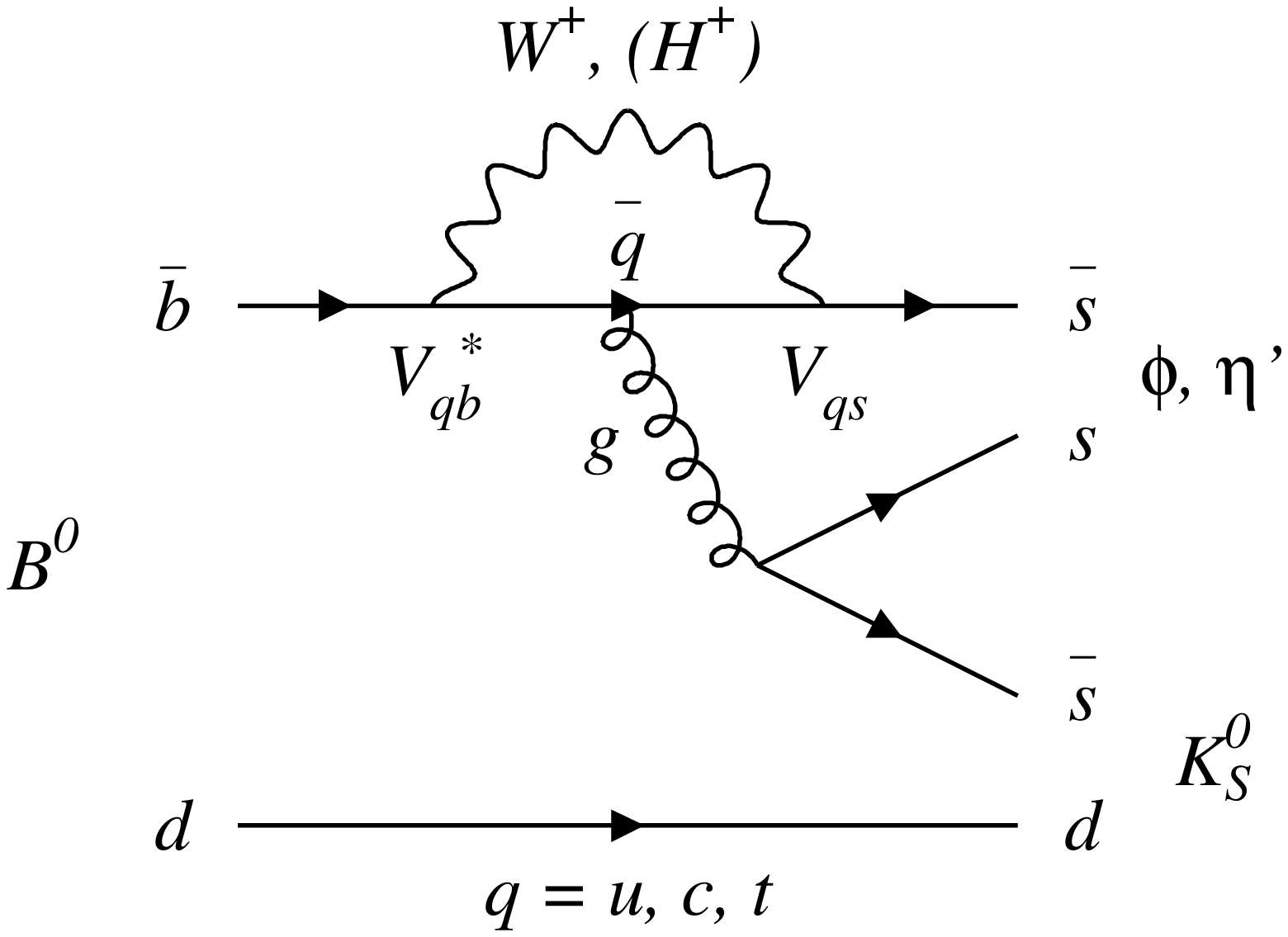}}
\vspace*{-0.8in}
{\caption [Eta' K_S final state diagram]
{\label{etaKdiagram}
The Feynman-diagram for the decay $B^0 \rightarrow \eta' K^0_S$.
The unexpectedly large branching fraction into this mode hints
that there may be extra contributions to the amplitude --
which could introduce additional $CP$
phases. }}
\end{center}
\end{figure}

Having looked at two of the more plentiful $B$ meson decays
that are $CP$ eigenstates, we turn to
the next most likely $B$ decay modes to show $CP$ phases.
One such class of decays is shown in fig.~\ref{etaKdiagram}.
With no tree level contribution these decay processes 
might be expected to have branching fractions 
smaller than those
discussed above. In fact, the mode $B \rightarrow \eta' K^0_S$
has a branching ratio that is somewhat larger than the mode
$B \rightarrow \pi^+\pi^-$ and clearly larger than the, as
yet unseen, decay $B \rightarrow \eta K$. That the $\eta' K$
branching ratio is so large (about $5 \times 10^{-5}$) is
not easy to understand theoretically -- hinting that there
may be more that just a charged $W$ boson involved in the
loop at the top of the diagram. If there were contributions 
from a
charged Higgs to this
amplitude, it could introduce an additional
phase, interfering with the weak phase predicted by the CKM model. In the 
absence of such additional phases this decay would provide another
way -- albeit a much lower statistics way -- to measure $\sin 2 \phi_1$.
Thus a comparison of the $CP$ asymmetry in the decay $\eta' K^0_S$
to that seen in $c {\bar c} K_S^0$ could help unravel the
mystery of the large branching of $B$ mesons into this mode
and may provide a glimpse of physics beyond the Standard Model.

\subsection{$CP$ Asymmetry in the Decay $B^0 \rightarrow \eta' K^0_S$}

\begin{figure}[ht]
\begin{center}
\vspace*{0in}
\mbox{\epsfxsize 7cm \epsfbox{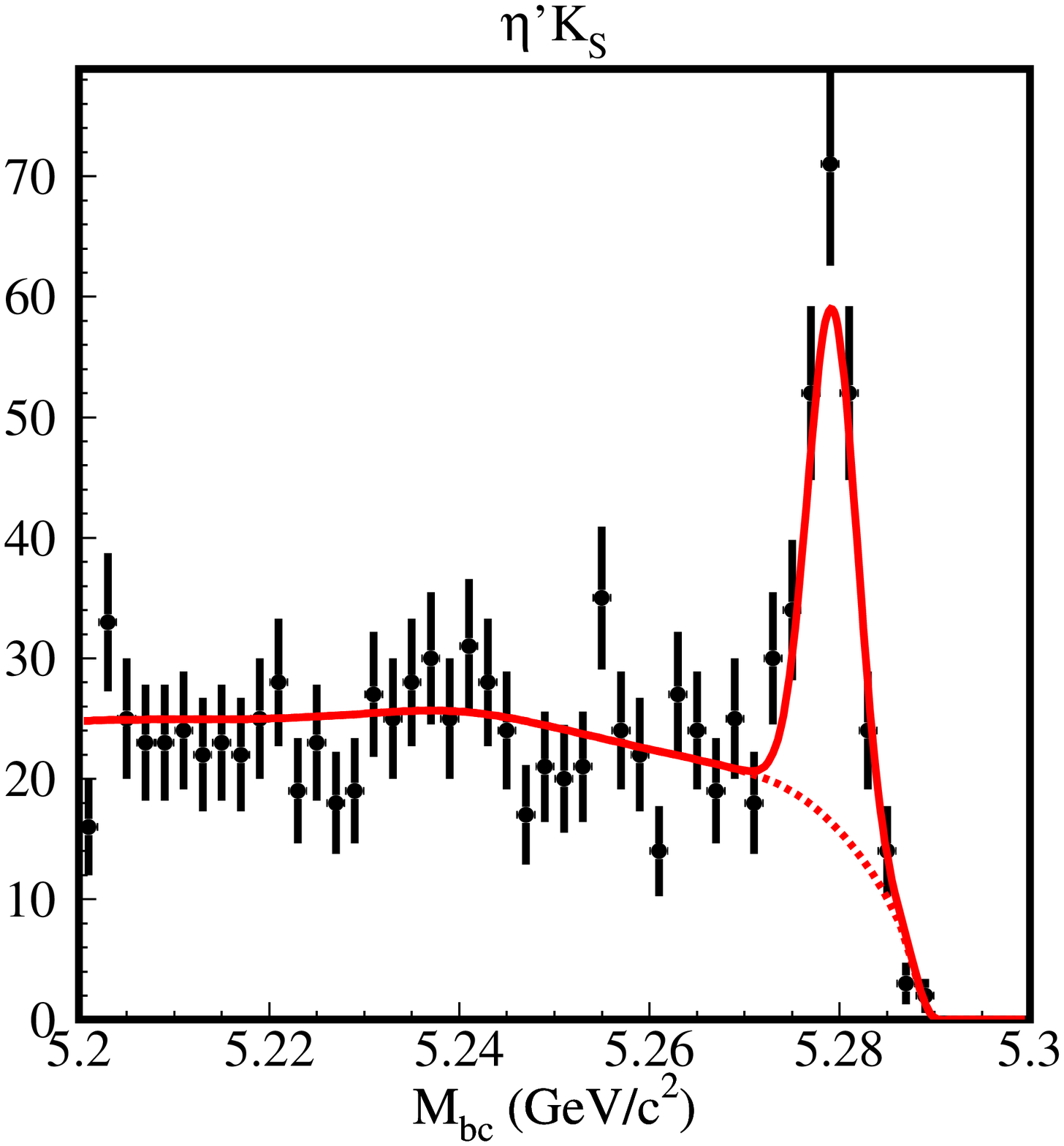}}
\mbox{\epsfxsize 7cm \epsfbox{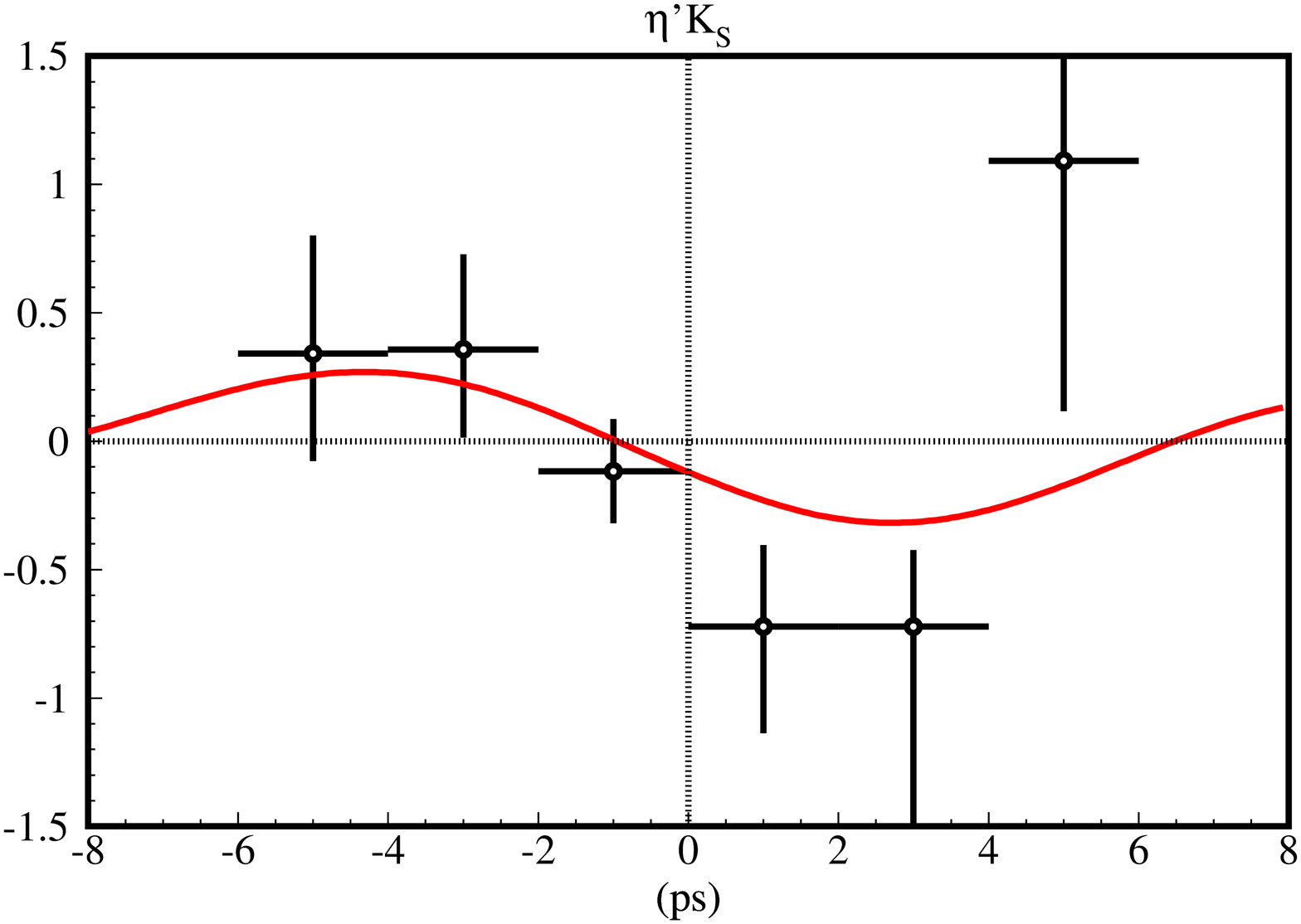}}
{\caption [Eta K_S signal]
{\label{etaks}
Left: The beam constrained mass distribution for
$B^0 \rightarrow \eta' K^0_S$ candidates. Right:
The $CP$ asymmetry as a function of $\Delta t$
for these candidates.
 }}
\end{center}
\end{figure}

Figure~\ref{etaks} shows the beam constrained
mass distribution for 128
candidate $B^0 \rightarrow \eta' K^0_S$ decays from 
78 fb$^{-1}$ of Belle data. Here the
$\eta'$ candidates have been reconstructed in $\eta \> \pi^+ \pi^-$
and $\rho^0 \gamma$ intermediate states and then combined
with $K_S^0 \rightarrow \pi^+\pi^-$
combinations. The plot on the right of fig.~\ref{etaks} 
shows the $B^0 {\bar B^0}$ asymmetry, as a function of
$\Delta t$, where the candidates have been flavour tagged
and vertexed 
in the same way as our $c {\bar c} K^0_S$ measurement of
$\sin 2 \phi_1$. The coefficient for indirect $CP$ violation
derived from this
fit (eqn.~\ref{asym_2} with $A \equiv 1$) is 
listed in table~\ref{tab:rare}. 
While the statistics are still poor, this is a demonstration
that we can measure $CP$ asymmetries in these rare modes.
More details on this analysis can be found in reference~\cite{etaks_plb}
that describes this measurement made on the first
half of the data shown here.

\renewcommand{\arraystretch}{1.1}
\begin{table}
\begin{center}
\begin{tabular}{l|cc}
\hline \hline
Mode                            & $(\sin 2 \phi_1 )_{\rm eff}$                                                     & Candidates \\ \hline
$\eta' K_S^0$                   & $ 0.76 \pm 0.36 ({\rm stat}) \pm 0.06 ({\rm sys})$                               &  128       \\
$\phi K_S^0$                    & $ -0.73 \pm 0.64 ({\rm stat}) \pm 0.18 ({\rm sys})$                              &  35        \\
$(K^+K^-)_{\rm non-res} K_S^0$  & $ 0.52 \pm 0.46 ({\rm stat}) \pm 0.11 ({\rm sys}) ^{+0.27} _{-0.03} ({\rm CP}) $ &  95        \\ \hline \hline
\end{tabular}
\end{center}
\caption{$CP$ fit parameters for $b \rightarrow s {\bar s} s$ decay modes.}
\label{tab:rare}
\end{table}

We perform the same analysis on samples of $B^0 \rightarrow K^+K^-K^0_S$
decays. Though the number of candidates in this sample is even
smaller, we can begin to probe their $CP$ asymmetry. We separate
the $K^+K^-$ candidates into two samples. When $m_{K^+K^-} \approx m_{\phi}$
the final state has $CP = -1$ providing another potentially
clean determination of $\sin 2 \phi_1$.  The statistics limited result
is shown in table~\ref{tab:rare}. The remaining sample is somewhat larger,
providing a more incisive measurement of the $CP$ asymmetry. Studies of
similar non-resonant decays related to this one by isospin 
(such as $B^+ \rightarrow K^+ K^0 {\bar K^0}$)
have shown that our $(K^+K^-)_{\rm non-res} K_S^0$ 
sample is $97^{+3}_{-16} \%$ $CP$ even.
This bound on the $CP$ symmetry of the non-resonant $K^+K^-$ final
state translates into an additional systematic on the effective $\sin 2 \phi$,
listed in table~\ref{tab:rare}. The measurement of
the $CP$ eigenvalue in this mode will improve with additional
data so the non-resonant mode might eventually provide 
another clean measurement of $\sin 2 \phi_1$, or a glimpse of a
non-Standard Model phase this decay.





\section{Summary and Future Prospects}

\begin{figure}
\begin{center}
\vspace*{-0.6in}
\mbox{\epsfxsize 14cm \epsfbox{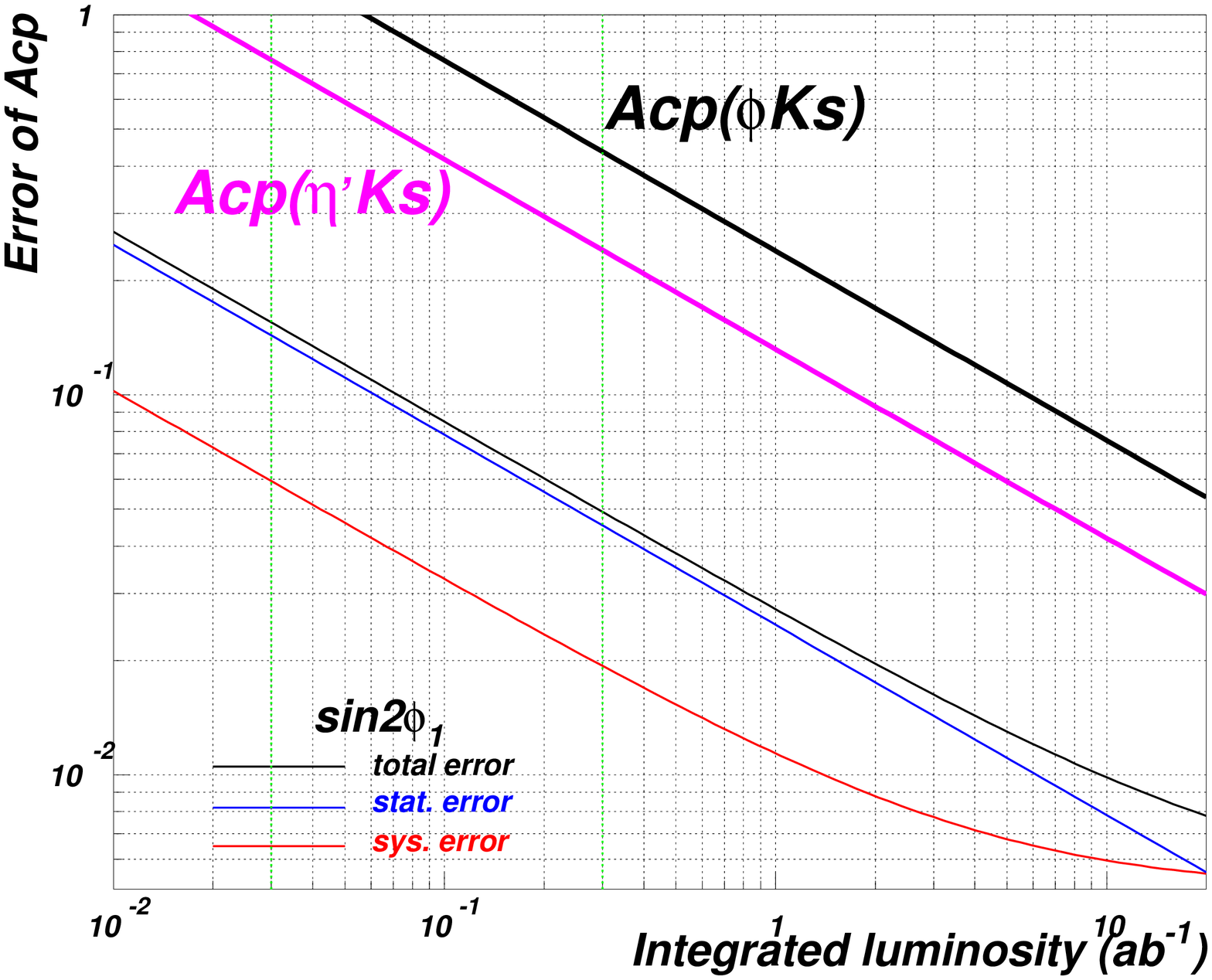}}
{\caption [Prospects for CP asymmetry precision]
{\label{cp_prospects}
Prospects for improving the precision on $\sin 2 \phi_1$
from the decay $c {\bar c} K^0_S$ (lower set of 
thin curves) with additional statistics. For
comparison the precision expected on rarer modes
($\eta' K_S^0$ thick grey curve and $\phi K^0_S$ thick
black curve) are also shown. }}
\end{center}
\end{figure}

Belle has studied the decay of $B^0$ mesons into
a number of $CP$ eigenstates. The most precise
measurement of $\sin 2 \phi_1$ to date comes from the modes
$B^0 \rightarrow c {\bar c} K$. Most of
the systematic uncertainties involved in that
measurement come from calibrations of our flavour
tagging and decay vertexing procedure based on control
samples. Thus it is reasonable to expect that the
precision on $\sin 2 \phi_1$ -- both statistical and systematic --
will improve as more data becomes
available. Figure~\ref{cp_prospects} shows a 
prediction of how the precision on $\sin 2 \phi_1$
will scale as Belle collects more data. The lowest curve shows the
evolution of the systematic uncertainty
on $\sin 2 \phi_1$ in the $c {\bar c} K$ measurement.
It should improve
until as least 1000 fb$^{-1}$ (1 ab$^{-1}$) of data has been collected.
At that point we expect an overall precision (uppermost thin line
in fig.~\ref{cp_prospects}) of
$\delta \sin 2 \phi_1 \approx 0.025$.

The rarer modes will also benefit from the
increase in statistics. Belle expects to be able to
accumulate 1 ab$^{-1}$ of data in the next 3 or 4 years.
At that point the precision on the $\eta' K^0_S$
asymmetry should approach the precision that we 
currently have on the $c {\bar c} K$ asymmetry. If there
are other modes contributing to the phase in $\eta' K^0_S$
decay un-natural fine-tuning would be necessary
for our precision not to 
reveal some discrepancy between what would otherwise
be two measurements of the same quantity.

The $B$ factories were built to confirm
the $CP$ violation predicted by the CKM model in
$B^0$ meson decay -- they have now achieved that milestone.
Having made a measurement of $\sin 2 \phi_1$ with
a precision of 10\% in
the $c {\bar c} K^0_S$ modes, and are now pushing
to improve the precision to a
few percent. At the same time they are expanding the
scope of their study to other decay processes. In doing
so they will measure other angles of the unitarity
triangle. This will over-constrain the CKM
model and test whether there is physics beyond the
three known families of quarks and their Standard
Model electroweak decays. In the coming years the
flavour sector will become ever more constrained
reducing the range of physics
that could explain why we have three families of quarks.

\end{document}